\documentclass[showpacs,aps,twocolumn]{revtex4}
\usepackage{epsfig}
\usepackage{graphicx}
\usepackage{amsmath,amssymb,amsfonts}
\usepackage{array}
\usepackage{url}
\usepackage{multirow}
\usepackage{float}
\usepackage{lineno}
\usepackage{xspace}
\usepackage{ulem}
\usepackage{bm}

\usepackage[usenames,dvipsnames]{color}
\definecolor{darkblue}{RGB}{0,0,196}
\usepackage[colorlinks=true,linkcolor=darkblue,citecolor=darkblue,urlcolor=darkblue]{hyperref}

\begin{document}
\title{Probing the microscopic origin of prompt and non-prompt $D^{0}$ production through event-shape engineering in proton-proton collisions at the LHC}
\author{Aswathy Menon Kavumpadikkal Radhakrishnan}
\author{Suraj Prasad}
\author{Purnima Srivastava}
\author{Raghunath Sahoo}\email[Corresponding Author: ]{Raghunath.Sahoo@cern.ch}
\affiliation{Department of Physics, Indian Institute of Technology Indore, Simrol, Indore 453552, India}

\begin{abstract}
Heavy-flavour hadrons are produced in the early stages of ultra-relativistic collisions at the LHC via hard partonic interactions and experience the whole system evolution. The study of prompt and non-prompt $D^{0}$ mesons provides an independent avenue to test the theories of quantum chromodynamics and to investigate beauty hadron production. Moreover, the production of both prompt and non-prompt $D^{0}$ is influenced by microscopic processes such as multi-partonic interactions (MPI) and hadronisation through fragmentation. In this study, an attempt is made to understand the production of prompt and non-prompt $D^{0}$ mesons in proton-proton collisions at $\sqrt{s}=13.6$ TeV using the PYTHIA8 event generator, which offers a qualitative description of charm production. The role of the transverse momentum transfer in the hardest partonic scattering ($\hat{p}_{\rm T}$), MPI, and color reconnection is systematically explored. In addition, the charged particle production in different topological regions with respect to the leading $D^{0}$ meson is studied to assess the influence of the $D^{0}$ meson on the event topology and to examine the selection biases arising from the use of charged particle multiplicity as an event classifier.
\end{abstract}

\date{\today}
\maketitle
\section{Introduction}
\label{sec:intro}
The Universe in its infancy is theorised to have been in a thermalised and deconfined state of strongly interacting partonic matter known as the quark–gluon plasma (QGP) phase. To investigate the properties of such matter and its subsequent evolution, two of the world's most powerful accelerators, namely, the Large Hadron Collider (LHC) at CERN, Switzerland, and the Relativistic Heavy-Ion Collider (RHIC) at BNL, USA, are employed to collide heavy ions at ultra-relativistic speeds. These nuclear collisions can create a transient medium of hot and dense nuclear matter which closely resembles the QGP formed in the early Universe~\cite{Yagi:2005yb}. However, due to its very short lifetime and due to the property of color confinement, this medium cannot be observed directly. Therefore, some indirect signatures are studied in heavy-ion collisions like jet quenching~\cite{Cunqueiro:2021wls}, quarkonia suppression~\cite{Matsui:1986dk}, strangeness enhancement~\cite{WA97:1999uwz}, etc., with proton-proton (pp) collisions as the baseline, where the formation of QGP is not anticipated. Surprisingly, the recent experimental measurements in high multiplicity pp collisions reveal features traditionally associated with heavy-ion collisions, such as strangeness enhancement~\cite{ALICE:2016fzo}, ridge-like structures~\cite{Li:2023vdj, OrtizVelasquez:2014jsg}, and indications of collective medium-like behaviour~\cite{CMS:2016fnw, Lim:2021spo}. 

To understand the QGP-like phenomena in high multiplicity pp collisions, different models are explored. In addition to the hydrodynamic models such as EPOS-LHC~\cite{Pierog:2013ria} and EPOS4~\cite{Werner:2023zvo}, the perturbative quantum chromodynamics (pQCD)-based models such as PYTHIA8~\cite{Sjostrand:2007gs} and HERWIG~\cite{Gieseke:2012ft} are able to mimic these QGP-like signatures in pp collisions. PYTHIA8 incorporates physical mechanisms such as multi-partonic interactions(MPI)~\cite{Sjostrand:2013cya}, color reconnection(CR)~\cite{OrtizVelasquez:2013ofg} and rope hadronisation~\cite{Bierlich:2014xba}, which explain most of the radial flow and strangeness enhancement features in pp collisions at the LHC energies~\cite{Prasad:2025yfj}. The QGP-like signals in pp collisions are enhanced for the events with a large number of MPI ($N_{\rm mpi}$)~\cite{OrtizVelasquez:2013ofg}, which is not an experimentally measurable quantity. Thus, in the conquest of understanding the QGP-like origin in pp collisions, the studies of event shape observables took a leap which indirectly probes the microscopic $N_{\rm mpi}$, thereby separating the hard-QCD dominated jetty events from those of isotropic events having large $N_{\rm mpi}$. At the LHC, several event-shape based studies use transverse sphericity~\cite{ALICE:2012cor}, transverse spherocity~\cite{Ortiz:2015ttf}, relative transverse activity classifier~\cite{Martin:2016igp}, charged particle flattenicity~\cite{Ortiz:2022zqr}, etc., to name a few~\cite{Prasad:2025yfj}. In particular, transverse spherocity has been extensively explored in both experimental measurements and in Monte Carlo (MC) event-generator frameworks to investigate QGP-like phenomena in small systems and to study multi-hadron production dynamics~\cite{MenonKavumpadikkalRadhakrishnan:2023cik, Prasad:2024gqq, Prasad:2025yfj}. Some of the observables include strangeness enhancement, particle ratios, partonic modification factor $Q_{\rm pp}$, kinetic freeze-out parameters extracted using radial flow velocity, etc.~\cite{Prasad:2024gqq, Khuntia:2018znt}. With this context, transverse spherocity ($S_{0}$) has been demonstrated to be a robust tool in selecting events in which the QGP-like effects are enhanced. Recent MC event generator-based studies further suggest that transverse spherocity can be effectively applied in heavy-ion collisions to select events with small or large elliptic flow and their fluctuations~\cite{Tripathy:2025npe}. Collectively, these findings highlight the versatility of transverse spherocity as an event-shape classifier across a range of collision systems, from small to large. For a recent review on event topology classifiers at the LHC era, see Ref. \cite{Prasad:2025yfj}.

That being said, the applicability of the event shape in the heavy-flavour sector is barely explored. However, the heavy-flavour hadrons are produced during earlier phases of the collisions and hence, influence the particle production.
Additionally, the study of heavy-flavour (HF) hadron production can provide crucial tests for models based on quantum chromodynamics (QCD). The heavy quarks, namely, the charm and beauty quarks, are formed during the initial stages of the collisions via large momentum transfers and witness the whole system's evolution. The calculations of inclusive production cross-sections of HF hadrons rely on factorisation schemes, where the $p_{\rm T}$-differential production cross-sections of HF hadrons include the convolution of the contributions from the parton distribution function of colliding nuclei, the partonic scattering cross-section and the fragmentation function, which parameterises the evolution of HF quark into a HF hadron~\cite{ALICE:2022tji}. The fragmentation functions are determined from $e^{+}e^{-}$ collisions~\cite{Gladilin:2014tba} assuming the hadronisation processes are ``universal" which does not depend upon collision energy and system~\cite{ALICE:2024oob}. However, the HF baryon-to-meson yield ratios at the LHC show a significant deviation from  $e^{+}e^{-}$ collisions, which indicates that the universality of hadronisation is not valid at the LHC energies~\cite{LHCb:2019fns, LHCb:2015qvk,ALICE:2023wbx, ALICE:2021dhb, Altmann:2024kwx, Minissale:2020bif, CMS:2019uws, ALICE:2020wfu, ALICE:2017thy, LHCb:2023wbo, ALICE:2022exq}. Here, the studies of prompt and non-prompt charm production can be useful to probe and understand distinct sectors of charm and beauty hadron production~\cite{ALICE:2022tji}. This includes, but is not limited to, different hadronisation processes that can describe the production of HF hadrons in collider experiments, which is a long-standing problem. 

The lightest open-charm hadron, the $D^{0}$ meson, can be produced either directly via charm-quark fragmentation or through the decay of higher-mass charm resonances, as well as from the decays of beauty hadrons. $D^{0}$ mesons produced directly in the initial hard scatterings or via decays of excited charm hadrons are classified as prompt. In contrast, $D^{0}$ mesons originating from the weak decays of beauty hadrons are referred to as non-prompt $D^{0}$. Measurements of prompt $D^{0}$ provide valuable insight into the formation of QCD medium and the transport properties of charm quark within it~\cite{Goswami:2022szb, Goswami:2023hdl}. Correspondingly, the non-prompt $D^0$ measurements offer important constraints on beauty hadron production mechanisms~\cite{ALICE:2022tji, ALICE:2023gjj, ALICE:2024xln}. In Pb--Pb collisions, the flow coefficients of prompt $D^{0}$ mesons are found to be larger than those of non-prompt $D^{0}$~\cite{ALICE:2023gjj, CMS:2022vfn}. This difference reflects distinct production dynamics and suggests a stronger interaction of charm hadrons within the QCD medium, possibly leading to partial thermalization. Additionally, finite elliptic flow has been observed for the prompt $D^0$ in pp and p-Pb collisions at low to intermediate $p_{\rm T}$ regions, while it is consistent with zero within uncertainties for non-prompt $D^0$~\cite{CMS:2020qul}. These observations indicate the fundamental differences in the production of prompt and non-prompt $D^0$ mesons, thereby motivating detailed investigations on their production in pp collisions. Additionally, experimental measurements show that the self-normalised yield of $D^0$ increases stronger-than-linearly with an increase in charged particle multiplicity~\cite{Giacalone:2021jdt, Bailung:2021jow}. This rise is even stronger for the non-prompt case than for prompt $D^0$ mesons and becomes more prominent with an increase in $p_{\rm T}$~\cite{Giacalone:2021jdt, Bailung:2021jow, Goswami:2024xrx}. These features of multiplicity-dependent $D^0$ production could be influenced by microscopic processes such as MPI and CR, as explored in Ref.~\cite{ALICE:2021kpy}, where the correlation between $D^{0}$ mesons and charged hadrons is studied. 

In this paper, pp collisions at $\sqrt{s}=13.6$ TeV are simulated using the event generator PYTHIA8, which provides a good qualitative description of charm production. Here, the production of $D^{0}$ is explored as a function of $N_{\rm mpi}$ and CR. Additionally, to complement the role of the experimentally inaccessible quantities --- MPI, and $\hat{p}_{\rm T}$, the momentum transfer in the hardest parton-parton interaction --- in $D^{0}$ production, transverse spherocity and $p_{\rm T}^{{\rm lead}-D^{0}}$ are employed as respective experimentally measurable proxies. The study also examines the role of the underlying event in $D^{0}$ production and the impact of hard-$D^{0}$ production on the final-state charged particle multiplicity.

The paper is organised as follows. The introduction and motivation are presented briefly in Section~\ref{sec:intro} followed by a discussion on event generation and methodology in Section~\ref{sec:method}. The results are presented and discussed in Section~\ref{sec:results}, and are finally summarised in Section~\ref{sec:summary}.

\section{Methodology}
\label{sec:method}
In this section, we present a brief introduction to event generation using PYTHIA8 and the event-shape classifier, transverse spherocity.

\subsection{PYTHIA8}
To test several phenomenological models and simulate various hadronic and heavy-ion collisions with known physical processes, Monte Carlo (MC) event generators are used.
PYTHIA8 is a state-of-the-art pQCD-based Monte-Carlo event generator which is often used to simulate hadronic, leptonic and heavy-ion collisions from RHIC to LHC energies. It includes soft and hard QCD processes and involves models for initial and final state parton showers, MPI, beam remnants, CR, string fragmentation, particle decays, etc.~\cite{PYTHIA8}. In this study, we have used PYTHIA8 (version 8.315), which is an upgrade over PYTHIA6 with a more realistic implementation of MPI, and explains the LHC results better. Here, the heavy-flavour quarks such as beauty and charm are produced through $2\rightarrow2$ hard sub-processes~\cite{Corke:2010yf}. The results reported in this paper are obtained by generating 5.2$\times$ $10^9$ minimum bias events for pp collisions at $\sqrt{s}$ = 13.6 TeV with processes 4C tune (\texttt{tune:pp=5}). Additionally, only non-diffractive and inelastic components of the total cross section for all the hard processes (\texttt{HardQCD:all=on}) are considered. To avoid the divergence of QCD processes in the limit $p_{\rm T}\xrightarrow{}0$, a cutoff of $\hat{p}_{\rm T}>0.5$~GeV/$c$ via \texttt{PhaseSpace:pTHatMinDiverge = 0.5} is implemented. We have used \texttt{ColourReconnection:mode = 2}. This mode refers to the gluon-move model, where the gluons are moved (or flipped) from one point to another in such a way that the string length is minimized. \texttt{Charmonium:all=on} and \texttt{Bottomonium:all=on} are used to enable all the charmonium and bottomonium production processes for the production of prompt and non-prompt $D^{0}$. We have stored all the $D^{0}$ mesons with their decay switched off (\texttt{421:mayDecay = false}). The event generation is repeated with MPI off and CR off to understand their impact on heavy-flavour hadron production.

\subsection{Charged particle multiplicity}
Unlike the heavy-ion collisions, where the impact parameter of the collisions is well defined, the impact parameter in point-like hadronic collisions is not trivially defined. Therefore, the centrality of collisions can not be defined in pp collisions. However, the charged particle multiplicity acts as a proxy for the centrality of collisions in heavy-ion collisions and can be used in proton-proton collisions. To classify one event from another, the charged particle multiplicity estimator is used. The estimation of the multiplicity estimator is performed in the forward rapidity region using the FT0 detector of the ALICE experiment at the LHC, known as FT0 Multiplicity or simply FT0M, whose acceptance region is  $-3.4<\eta<-2.3$ and $3.8<\eta<5.0$. The definition of the multiplicity classes and corresponding percentile cuts of FT0M from PYTHIA8-generated events are shown in Table~\ref{Table:1}.
 
\subsection{Transverse Spherocity}
Transverse spherocity ($S_{0}$) is an event shape observable that is capable of separating events based on the geometrical distribution of the final state particles produced in hadronic and nuclear collisions. It is calculated using the following expression~\cite{Cuautle:2014yda, Ortiz:2015ttf, Ortiz:2017jho, Prasad:2025yfj, ALICE:2023bga}.
\begin{equation}
    S_{0} =\frac{\pi ^2}{4}\min_{\hat{n}}\left(\frac{\sum_{i=1}^{N_{\rm ch}}|\hat{p}_{{\rm T}_{i}}\times {\hat{n}}|}{N_{\rm ch}}\right)^2
    \label{eq:1S0pT1defn}
\end{equation}
Here, $\hat{p}_{\rm T_{i}}$ is a unit vector along the direction of the transverse momentum of $i^{\rm th}$ charged particle. $i$ runs over all the final-state charged particles, and $N_{\rm ch}$ is the total number of charged particles in an event. $\hat{n}(n_{\rm T},0) $ is the unit vector which is chosen in such a way that it minimises the ratio presented in parentheses in Eq.~\eqref{eq:1S0pT1defn}.  Multiplication of $\pi^2/4$ makes sure that the range of $S_{0}$ is between 0 and 1. The lower limit ($S_{0}\rightarrow0$) corresponds to ``jetty events" with back-to-back (pencil-like) emission of particles dominated by hard QCD processes, whereas the higher limit ($S_{0}\rightarrow1$) represents ``isotropic events" which is dominated with soft QCD interaction and the particle emission is thus uniform in azimuth in the transverse plane. By definition, $S_{0}$ is infrared and collinear safe~\cite{Salam:2010nqg, Prasad:2025yfj}. 

In this study, the events with the lowest 20$\%$ values of $S_0$ are referred to as jetty events, while the highest 20$\%$ in the $S_{0}$ distribution are considered to be the isotropic events. Furthermore, only the events having more than 10 charged particles in the pseudorapidity range of $|\eta|<0.8$ with $|p_{\rm T}|>0.15$~GeV/$c$ are considered for this study~\cite{Prasad:2025yfj}, referred to as the $S_0$-integrated events. Table~\ref{Table:1} contains the spherocity cuts for jetty and isotropic events in each FT0M class, for both the CR-on--MPI-on ((a)) and the CR-off--MPI-on ((b)) cases. 

\begin{table}[h]
\caption{\label{spherocutPYTHIA} Transverse spherocity cuts for the jetty and isotropic events for different multiplicity classes with the corresponding charged particle multiplicities in mid rapidity in pp collisions at $\sqrt{s}$ = 13.6 TeV using PYTHIA8 for: \label{Table:1}} 
   (a) CR-on and MPI-on case
\begin{tabular}{|c| c| c| c| c| c|}
\hline 
\multirow{2}{*}{{\shortstack{FT0 \\ Percentile}}} & \multirow{2}{*}{{FT0M}} & \multirow{2}{*}{\bf{$\langle dN_{\rm ch}/d\eta \rangle_{|\eta|<0.8}$}} & \multicolumn{2}{c|}{\bf{$S_{0}$($|\eta|<0.8$)}} \\\cline{4-5}     & &&  {{Jetty}}  & {{Isotropic}}  \\ \hline
0--1 &46-200 &26.8715 $\pm$ 0.0054  & 0 --0.689   & 0.864 -- 1       \\ \hline
1--5 &26-46   & 15.8879 $\pm$ 0.0021 & 0 -- 0.614   & 0.830 -- 1       \\ \hline
5--10 &18-26  &9.0967 $\pm$ 0.0012 & 0 -- 0.558  & 0.800 -- 1       \\ \hline
10--20 &13-18 &5.6913 $\pm$ 0.0007  & 0 -- 0.531& 0.784 -- 1       \\ \hline
20--40 &9-13  &3.6674 $\pm$ 0.0003  & 0 --  0.514  & 0.775 -- 1       \\ \hline
40--60 &7-9    & 2.9456 $\pm$ 0.0003  & 0 -- 0.503   & 0.768 -- 1       \\ \hline
60--90 &3-7  &2.6348 $\pm$ 0.0002  & 0 -- 0.491   &0.762 -- 1       \\ \hline  \hline
0--100 &0-200 &4.3280 $\pm$ 0.0003   & 0 -- 0.557  & 0.807 -- 1 \\ 
\hline
\end{tabular}

   (b) CR-off and MPI-on case
\begin{tabular}{|c| c| c| c| c| c|}
\hline 
\multirow{2}{*}{{\shortstack{FT0 \\ Percentile}}} & \multirow{2}{*}{{FT0M}} & \multirow{2}{*}{\bf{$\langle dN_{\rm ch}/d\eta \rangle_{|\eta|<0.8}$}} & \multicolumn{2}{c|}{\bf{$S_{0}$($|\eta|<0.8$)}} \\\cline{4-5}     & &&  {{Jetty}}  & {{Isotropic}}  \\ \hline
0--1 &58-200  &34.1083 $\pm$ 0.0068  & 0 --0.722   & 0.879 -- 1       \\ \hline
1--5 &32-58   & 19.7527 $\pm$ 0.0026 & 0 -- 0.644   & 0.845-- 1       \\ \hline
5--10 &22-32  &11.3114 $\pm$ 0.0016 & 0 -- 0.572  & 0.808 -- 1       \\ \hline
10--20 &15-22 &7.0154 $\pm$ 0.0008   & 0 -- 0.533& 0.786 -- 1       \\ \hline
20--40 &10-15  &4.3060 $\pm$ 0.0004  & 0 --  0.510  & 0.772 -- 1       \\ \hline
40--60 &7-10    & 3.1318 $\pm$0.0003  & 0 -- 0.495   & 0.763 -- 1       \\ \hline
60--90 &4-7    &2.7440 $\pm$ 0.0002 & 0 -- 0.483   &0.756 -- 1       \\ \hline \hline
0--100 &0-200 &5.0516 $\pm$0.0004   & 0 -- 0.552  & 0.807 -- 1 \\ 
\hline
\end{tabular}
\end{table}

\section{Results and discussions}
\label{sec:results}

\begin{figure}[ht!]
\begin{center}
\includegraphics[scale = 0.42]{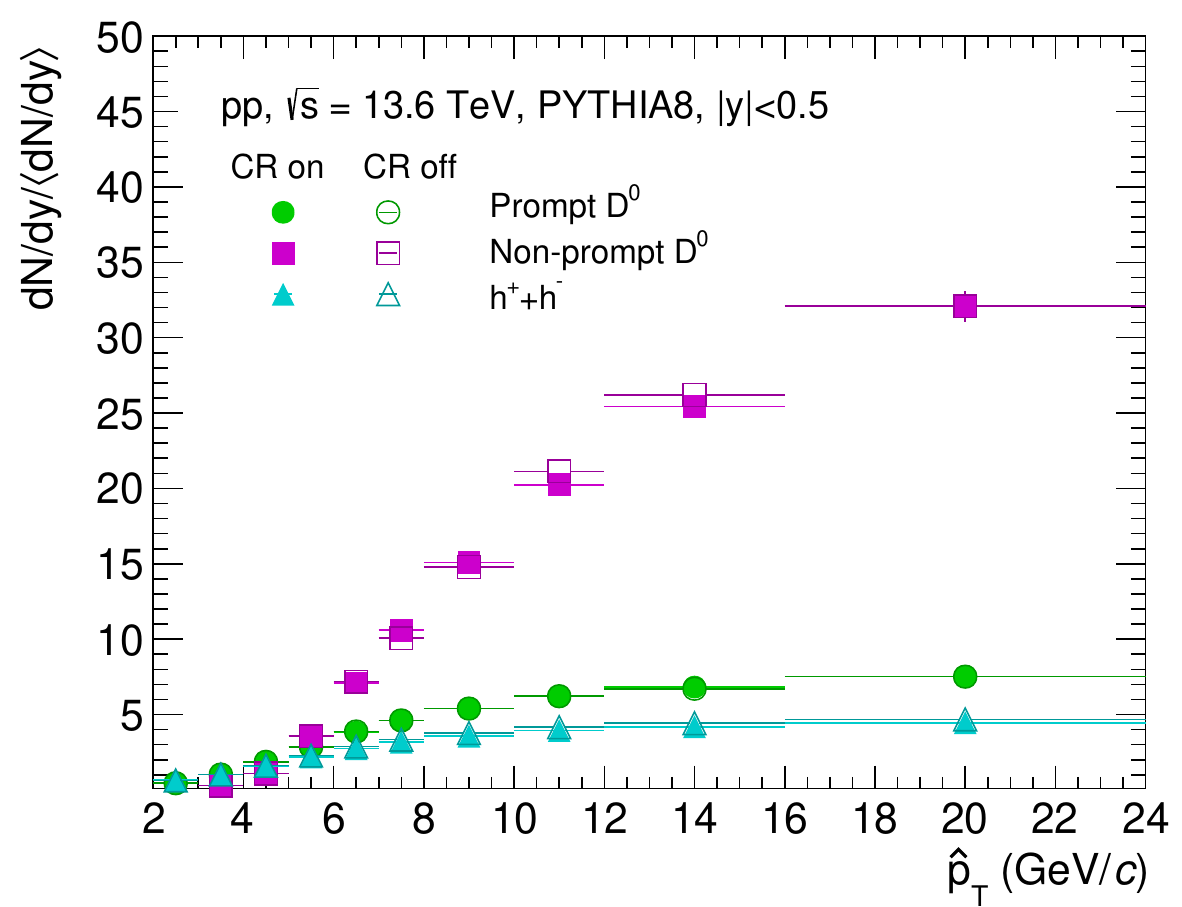}
\caption{Self-normalised yield of prompt and non-prompt $D^{0}$ measured at mid-rapidity compared with that of charged particles as a function of $\hat{p}_{\rm T}$ in pp collisions at $\sqrt{s}$ = 13.6~TeV using PYTHIA8.}
\label{fig:D0vspThat}
\end{center}
\end{figure}

\begin{figure}[ht!]
\begin{center}
\includegraphics[scale = 0.42]{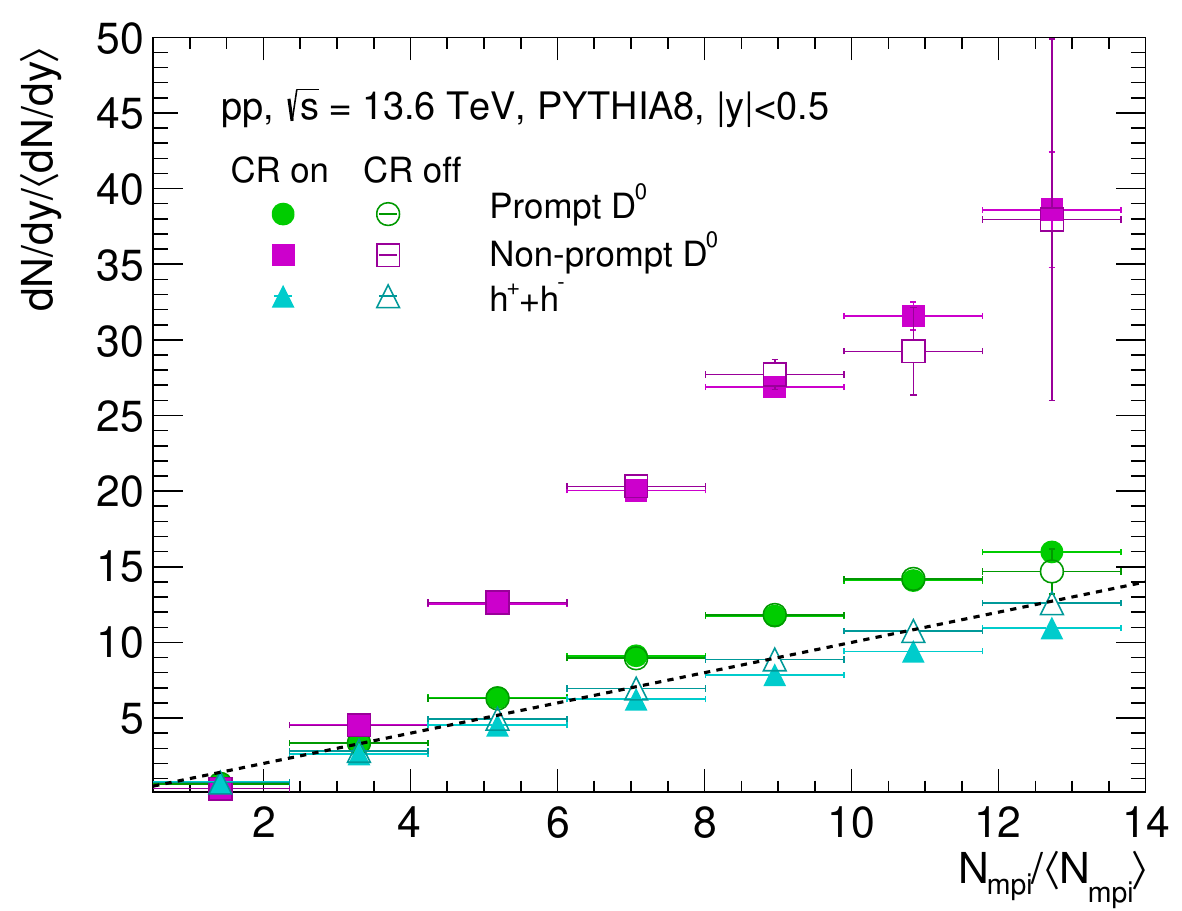}
\caption{Self-normalised yield of prompt and non-prompt $D^{0}$ measured at mid-rapidity compared with that of charged particles as a function of $N_{\rm mpi}$ in pp collisions at $\sqrt{s}$ = 13.6 TeV using PYTHIA8.}
\label{fig:D0vsNmpi}
\end{center}
\end{figure}

This section starts with the demonstration of production of prompt and non-prompt $D^0$ and charged particles as a function of $\hat{p}_{\rm T}$ and $N_{\rm mpi}$, shown in Fig~\ref{fig:D0vspThat} and~\ref{fig:D0vsNmpi}, respectively. 

\textit{Particle production and $\hat{p}_{\rm T}$}: 
Figure~\ref{fig:D0vspThat} shows the self-normalised yield of prompt $D^{0}$, nonprompt $D^{0}$ and all charged particles plotted as function of transverse momentum transfer of the hardest parton–parton interaction ($\hat{p}_{\rm T}$) in pp collision at $\sqrt{s} = 13.6$ TeV using PYTHIA8. The results for the CR-on case are compared with the CR-off case, and the difference between them is found to be negligible. As a function of $\hat{p}_{\rm T}$, the increase in the self-normalised yield is the largest for non-prompt $D^{0}$ mesons, followed by prompt $D^{0}$ mesons, and is the smallest for all-charged hadrons. This hierarchy reflects a mass ordering of the particles under consideration: charged hadrons (dominated by light mesons such as pions) exhibit the weakest rise, whereas prompt and non-prompt $D^{0}$ mesons show a stronger increase with $\hat{p}_{\rm T}$. This is because the production probability of massive particles increases rapidly with the momentum transfer in the hardest parton-parton interactions, where the heavy-flavour quarks are predominantly produced. In contrast, the charged particles can originate from multiple stages of space-time evolution of proton-proton collisions through a variety of soft and semi-hard processes, making their production less sensitive to $\hat{p}_{\rm T}$. However, one observes a pronounced difference in the increase of self-normalised yield of prompt and nonprompt $D^{0}$ with $\hat{p}_{\rm T}$. This indicates that, unlike non-prompt $D^{0}$, prompt $D^{0}$ production receives substantial contributions from mechanisms beyond the hardest parton–parton scatterings.

\textit{Particle production and $N_{\rm mpi}$}: Before we proceed to understand the particle production as a function of $N_{\rm mpi}$, it is useful to recall the correlation between the average momentum transfer in the hardest parton–parton interaction, $\langle\hat{p}_{\rm T}\rangle$ and the average number of multiparton interactions, $\langle N_{\rm mpi}\rangle$ for events selected based on $N_{\rm mpi}$, as previously discussed in Refs.~\cite{Prasad:2025yfj, Prasad:2024gqq}. Here, $\langle\hat{p}_{\rm T}\rangle$ increases with an increase in $\langle N_{\rm mpi}\rangle$ and shows a saturating trend towards higher $\langle N_{\rm mpi}\rangle$ events. Note that, in the framework of PYTHIA8, the $N_{\rm mpi}$ depends on the impact parameter ($b$) of the collision. As $b$ decreases, the overlap between the colliding protons increases, leading to a larger number of overlapping strings and, consequently, to higher values of $N_{\rm mpi}$ and $\hat{p}_{\rm T}$. For instance, events with $b\simeq0$ are expected to develop many overlapping strings during collision, resulting in larger $\langle\hat{p}_{\rm T}\rangle$ and $\langle N_{\rm mpi}\rangle$ than events with higher values of $b$. Moreover, MPIs occur after the initial hard scattering and are restricted by the total energy available for particle production, thereby leading to a saturation of $\langle\hat{p}_{\rm T}\rangle$ towards higher $\langle N_{\rm mpi}\rangle$. 

This relation between $\hat{p}_{\rm T}$ and $N_{\rm mpi}$ is reflected in the production of non-prompt $D^0$ when studied as a function of $N_{\rm mpi}$. The probability of producing non-prompt $D^0$ (or beauty hadron) increases steeply with the number of strings undergoing large transverse momentum transfer, which becomes more likely as $b$ approaches zero and $N_{\rm mpi}$ increases. This is reflected in the relative yield of non-prompt $D^0$ as a function of relative $N_{\rm mpi}$, shown in Fig.~\ref{fig:D0vsNmpi}. On the other hand, the charm quarks can be produced at later stages of the collisions, involving smaller transverse momentum transfer, and thus, the relative increase in the production of prompt $D^{0}$ with the relative $N_{\rm mpi}$ is weaker than that of the non-prompt $D^0$. Finally, the relative production of charged particles, which are produced in different stages of the collision, scales linearly with $N_{\rm mpi}/\langle N_{\rm mpi}\rangle$ for the CR-off case and exhibits a weaker-than-linear scaling for the CR-on case, as evident from Fig.~\ref{fig:D0vsNmpi}. It is interesting to note that the relative production of prompt and non-prompt $D^0$ depends only weakly on CR, which occurs at a later stage in the evolution of the collision.

\begin{figure}[ht!]
\includegraphics[scale = 0.44]{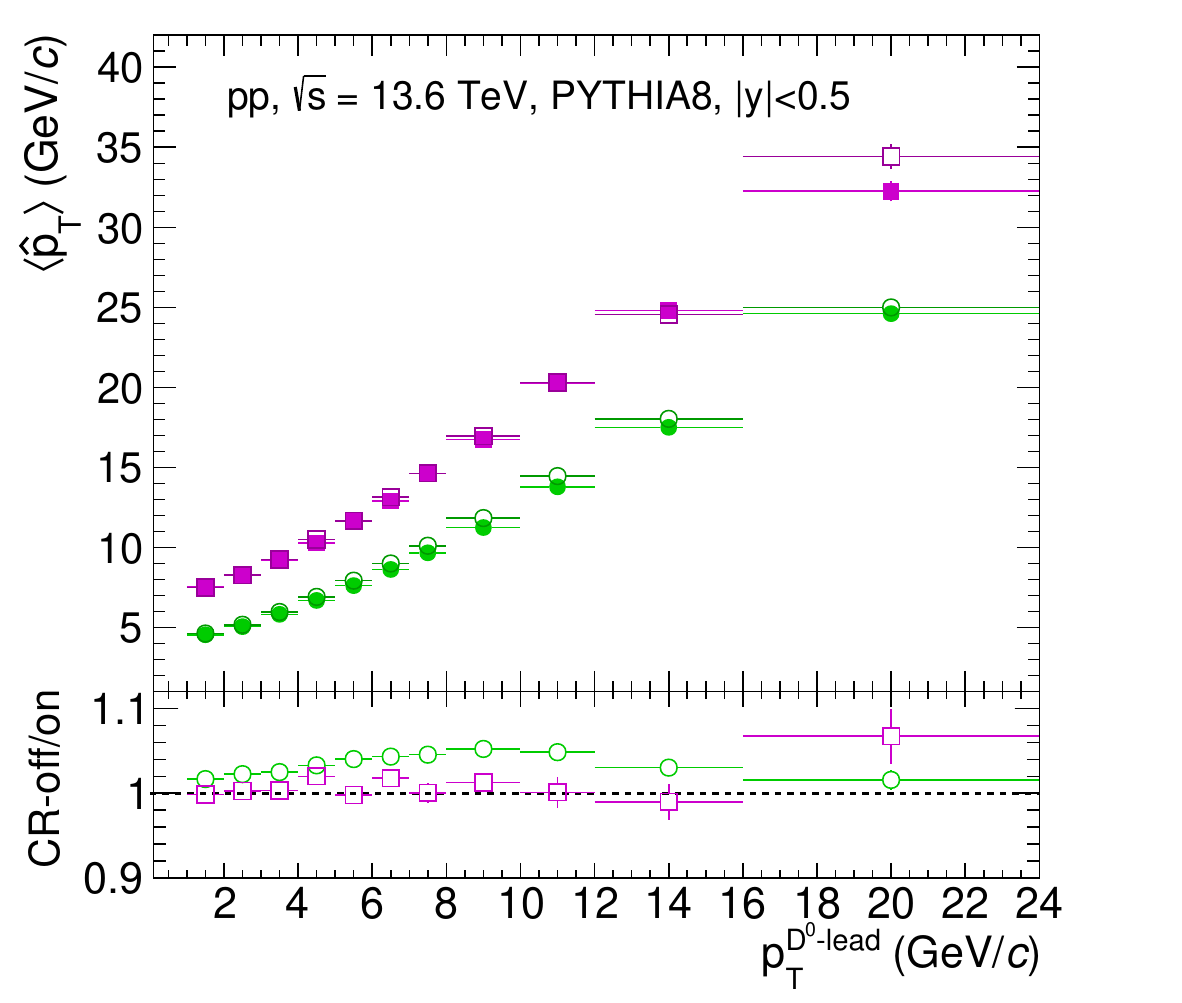}
\includegraphics[scale = 0.44]{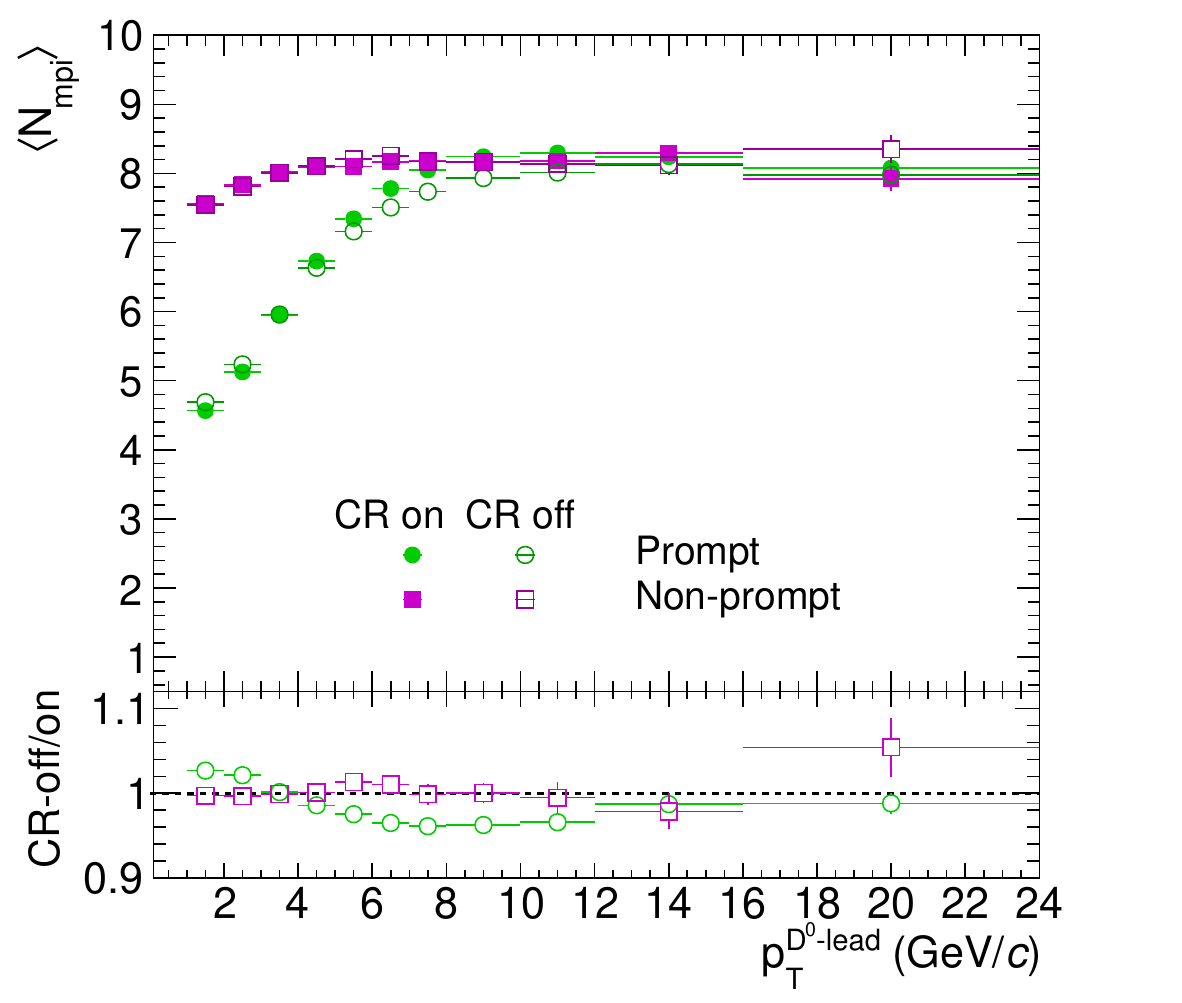}
\caption{$\langle\hat{p}_{\rm T}\rangle$ (top) and  $\langle N_{\rm mpi}\rangle$ (bottom) as function of $p_{\rm T}^{D^{0}-\rm lead}$ for prompt and non-prompt $D^{0}$ mesons in  pp collisions at $\sqrt{s} = 13.6$ TeV measured at midrapidity ($|y|<0.5$) using PYTHIA8. The lower panel in both plots illustrates the CR off-to-CR on ratio. }
\label{fig:pTd0leadvspThat}
\end{figure}

Although the production of prompt and non-prompt $D^0$ is sensitive to $\hat{p}_{\rm T}$ and $N_{\rm mpi}$ of an event, they are experimentally not measurable. Instead, one can look for alternate observables which are sensitive to $N_{\rm mpi}$ and $\hat{p}_{\rm T}$ and provide a foundation for understanding the HF production mechanism. As shown in the upper plot of Fig.~\ref{fig:pTd0leadvspThat}, the transverse momentum of leading $D^0$ ($p_{\rm T}^{D^{0}-{\rm lead}}$) is found to be well correlated with $\hat{p}_{\rm T}$, where the $p_{\rm T}^{D^{0}-{\rm lead}}$ for non-prompt covers a region in $\langle\hat{p}_{\rm T}\rangle$ which is higher than that of prompt $D^0$. This could be attributed to the requirement of a higher $\langle \hat{p}_{\rm T}\rangle$ to produce a non-prompt $D^{0}$ than a prompt $D^0$. It is noteworthy that $\langle\hat{p}_{\rm T}\rangle$ has slight dependence on color reconnection for the prompt-$D^0$ case, which is absent within uncertainties for the non-prompt $D^0$ case. $\langle\hat{p}_{\rm T}\rangle$ of prompt $D^0$ for CR-off case is slightly (less than 5\%) higher than that of CR-on case. This is because, in CR-on case, the strings arising from independent partonic scatterings can color reconnect which has a higher probability of weakening the transformation of the initial $\hat{p}_{\rm T}$ to $p_{\rm T}^{D^{0}-{\rm lead}}$ than the CR-off case where the partons hadronize independently from individual partonic interactions. The independence of $\langle\hat{p}_{\rm T}\rangle$ as a function of $p_{\rm T}^{D^{0}-{\rm lead}}$ on CR conditions for the non-prompt $D^0$ case once again indicates that the beauty quark production is predominantly governed by early hard scatterings and is therefore, weakly affected by the late-stage color-reconnection effects.

Moreover, in the lower plot of Fig.~\ref{fig:pTd0leadvspThat}, which shows $\langle N_{\rm mpi}\rangle$ versus $p_{\rm T}^{D^{0}-\rm lead}$ for prompt and non-prompt $D^{0}$ mesons, a qualitative similarity is observed with $N_{\rm mpi}$ versus $\hat{p}_{\rm T}$. This qualitative similarity is visually stronger for prompt $D^0$ than non-prompt $D^0$, as the latter requires events with a small value of $b$, which inherently have large $N_{\rm mpi}$. In contrast, the prompt $D^0$ does not necessitate such conditions, and thus covers a broad range of $\langle N_{\rm mpi}\rangle$. Nevertheless, to produce a charm quark with large momenta, the events need to have large enough energy, the probability of which increases with a decrease in $b$. As a consequence, $\langle N_{\rm mpi}\rangle$ increases with $p_{\rm T}^{D^{0}\text{-}\rm lead}$ and saturates for $p_{\rm T}^{D^{0}\text{-}\rm lead} \gtrsim 10$~GeV/$c$. This saturation arises from the finite amount of energy available in a pp collision, which limits the total number of MPIs that can occur~\cite{Weber:2018ddv}.

Similar to the upper plot for $\langle \hat{p}_{\rm T}\rangle$, in the lower plot, one does not see a significant difference in $\langle N_{\rm mpi}\rangle$ between CR-on and CR-off cases for the non-prompt $D^0$ mesons. However, a crossing in $\langle N_{\rm mpi}\rangle$ is observed between CR-on and off cases for the prompt-$D^0$ meson, where the $\langle N_{\rm mpi}\rangle$ is higher for CR-off than CR-on in $p_{\rm T}^{D^{0}-\rm lead}\lesssim 4$~GeV/$c$ region with the trend for CR-on case starting to dominate after $p_{\rm T}^{D^{0}-\rm lead}\gtrsim 4$~GeV/$c$. In other words, to produce a prompt $D^0$ with $p_{\rm T}^{D^{0}-\rm lead}\lesssim 4$~GeV/$c$, an event with a slightly higher $\langle N_{\rm mpi}\rangle$ is necessary when CR is switched off, in contrast to when CR is turned on. Contrary to this, the events with CR-on require larger $\langle N_{\rm mpi}\rangle$ than CR-off to produce prompt-$D^0$ with $p_{\rm T}^{D^{0}-\rm lead}\gtrsim 4$. This can be understood as follows:
\begin{itemize}
    \item $p_{\rm T}^{D^{0}\text{-}\rm lead}\lesssim 4$~GeV/$c$: In the CR-on scenario, color reconnection allows the redistribution of $p_{\rm T}$ among partons originating from different partonic scatterings. As a result, a charm quark produced in one partonic interaction can form a leading prompt $D^0$ meson in the low-$p_{\rm T}$ region, even in events with relatively fewer MPIs. In contrast, in the absence of color reconnection when all partonic scatterings remain independent, producing a prompt $D^{0}$ with similar $p_{\rm T}$ requires events with larger partonic overlap, which naturally leads to higher MPI activity.

    \item $p_{\rm T}^{D^{0}\text{-}\rm lead}\gtrsim 4$~GeV/$c$: 
    In the high-$p_{\rm T}$ regime, the production of a leading prompt $D^{0}$ is dominated by a hard parton-parton scattering. In the CR-off scenario, the occurrence of one such hard interaction is sufficient, and the additional MPIs do not significantly enhance the probability of producing a leading $D^{0}$. When color reconnection is enabled, the multiple color strings present in high-MPI events can rearrange during hadronisation, which can influence how the momentum of an already-produced charm quark is distributed among final-state hadrons. This string-level momentum redistribution can slightly modify the probability of forming a leading prompt $D^{0}$ with a given high transverse momentum.
\end{itemize}

\begin{figure}[ht!]
\begin{center}
\includegraphics[scale = 0.44]{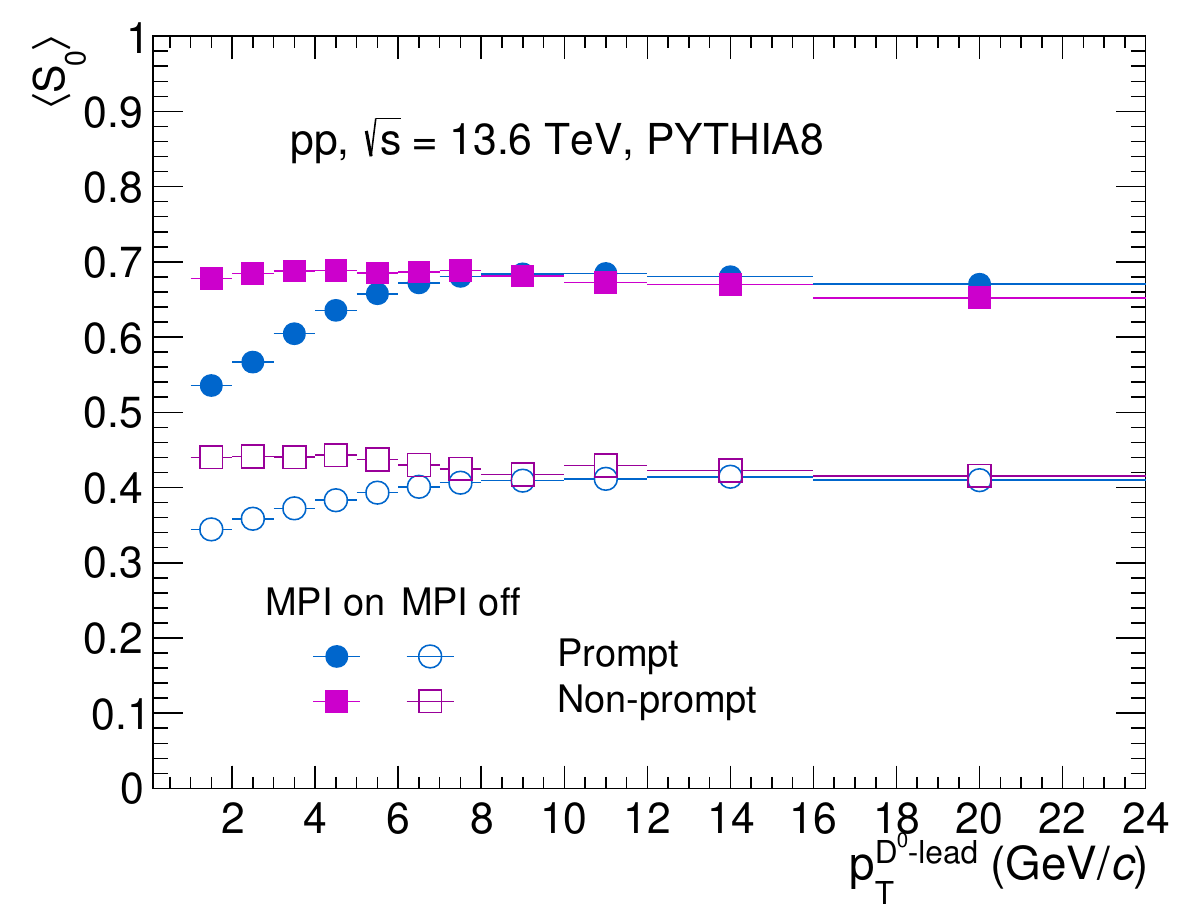}
\caption{$\langle S_{0}\rangle$ versus $p_{\rm T}^{D^{0}-\rm lead}$ for MPI-on and MPI-off cases for prompt and non-prompt $D^{0}$ mesons in pp collisions at $\sqrt{s}~=~13.6$~TeV using PYTHIA8 for CR-on case.}
\label{fig:pTD0leadvsS0}
\end{center}
\end{figure}

This correlation of $\langle N_{\rm mpi}\rangle$ and $p_{\rm T}^{D^0-\rm lead}$ is qualitatively reproduced and shown in Fig.~\ref{fig:pTD0leadvsS0}, when $N_{\rm mpi}$ is replaced with transverse spherocity ($S_0$). Here, for prompt $D^{0}$, the events with large $p_{\rm T}^{D^0-\rm lead}$ have larger $\langle S_0\rangle$ value following a saturation behaviour of $\langle S_0\rangle$ for $p_{\rm T}^{D^0-\rm lead}\geq 8$~GeV/$c$. Moreover, for non-prompt $D^0$, irrespective of $p_{\rm T}^{D^0-\rm lead}$, the corresponding $\langle S_0\rangle$ remains fairly constant, which indicates that the non-prompt $D^0$ production is weakly correlated with the global event shape and is primarily driven by the hardest partonic scattering rather than by MPI-induced isotropy. When $N_{\rm mpi}$ is set to 1, the average value of $S_0$ drops drastically for both prompt and non-prompt $D^0$ cases, revealing the sensitivity and usefulness of $S_0$ to probe microscopic $N_{\rm mpi}$. Hence, in subsequent results, the observables are shown as a function of $p_{\rm T}^{D^0-\rm lead}$ and $S_0$.

\begin{figure*}[ht!]
\begin{center}
\includegraphics[scale = 0.42]{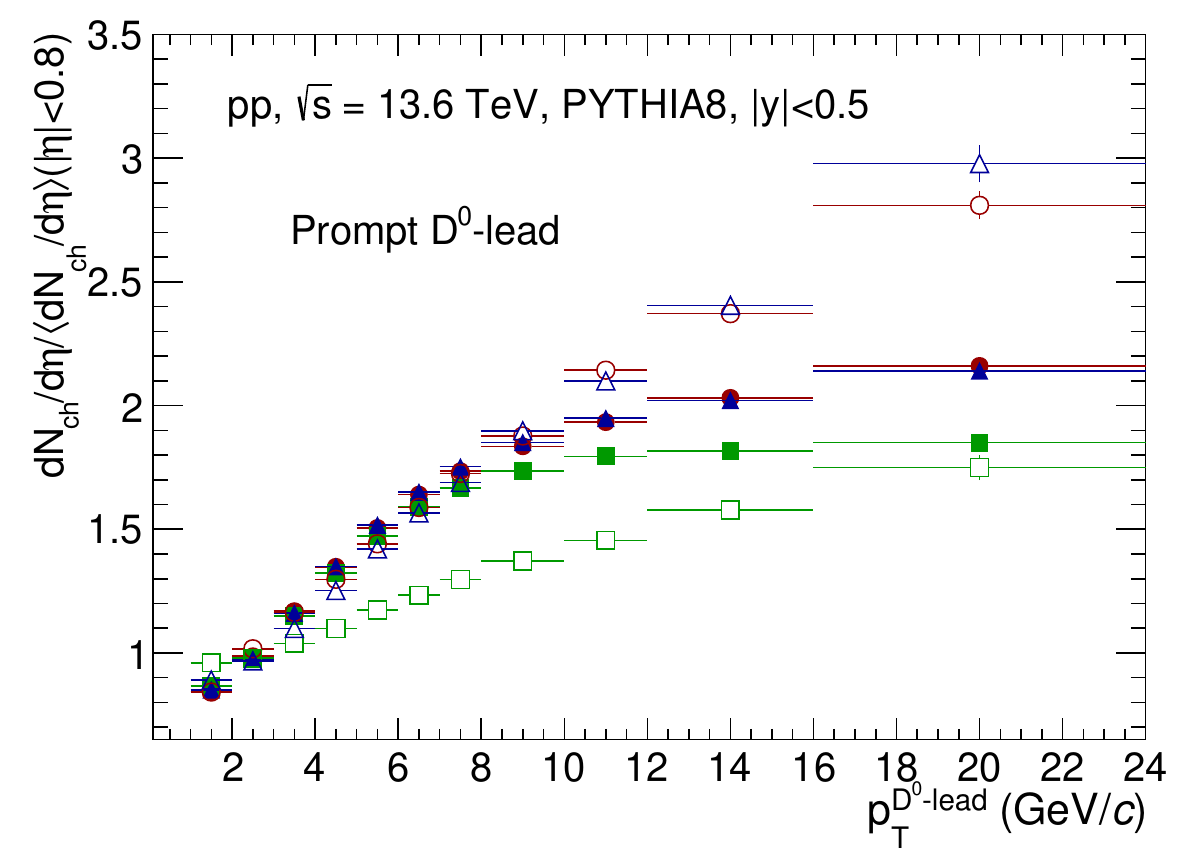}
\includegraphics[scale = 0.42]{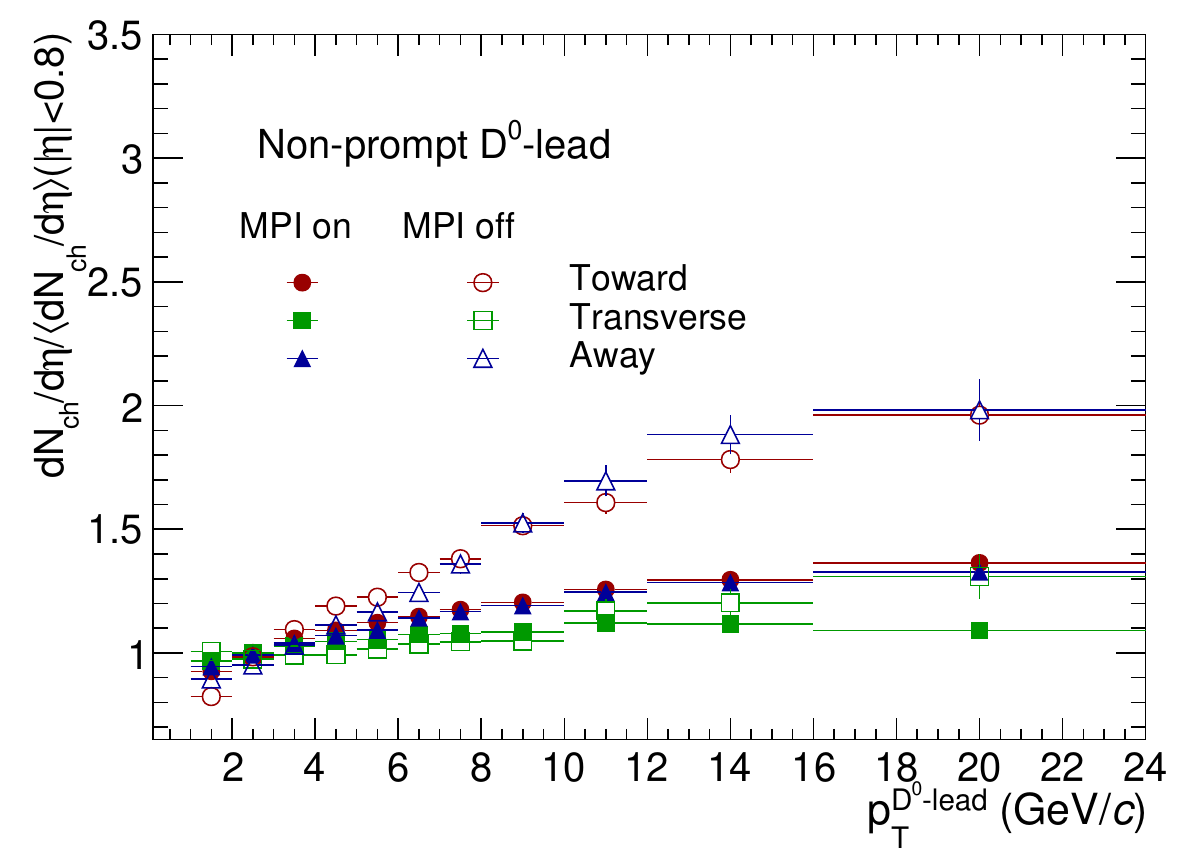}
\caption{Self-normalised mid-rapidity charged particle multiplicity density versus $p_{\rm T}^{D^{0}-\rm lead}$ for toward, transverse and away regions defined with respect to the leading prompt and non-prompt $D^{0}$ for ``MPI on" and ``MPI off" cases in pp collisions at $\sqrt{s} = 13.6$ TeV using PYTHIA8.}
\label{fig:pTD0leadvsD0region}
\end{center}
\end{figure*}

Further insights into the production mechanism of $D^{0}$ can be obtained by dividing the azimuthal plane into three topological regions relative to the azimuthal angle ($\Delta \phi = |\phi - \phi_{D^{0}}|$) of the leading $D^{0}$ meson i.e. the highest-$p_{\rm T}$ $D^{0}$ meson in an event, as follows.
\begin{enumerate}
    \item Toward region: $\Delta{\phi}\leq\pi/3$
    \item Transverse region: $\pi/3<\Delta{\phi}\leq2\pi/3$
    \item Away region: $\Delta{\phi}>2\pi/3$
\end{enumerate}
Here, toward and away regions show the particle production along and opposite to the azimuthal direction of the leading $D^0$, and are expected to be highly influenced by the production of $D^0$. The transverse region, on the other hand, is expected to be least affected by the physical mechanisms that involve the production of $D^0$. 

\begin{figure*}[ht!]
\begin{center}
\includegraphics[scale = 0.4]{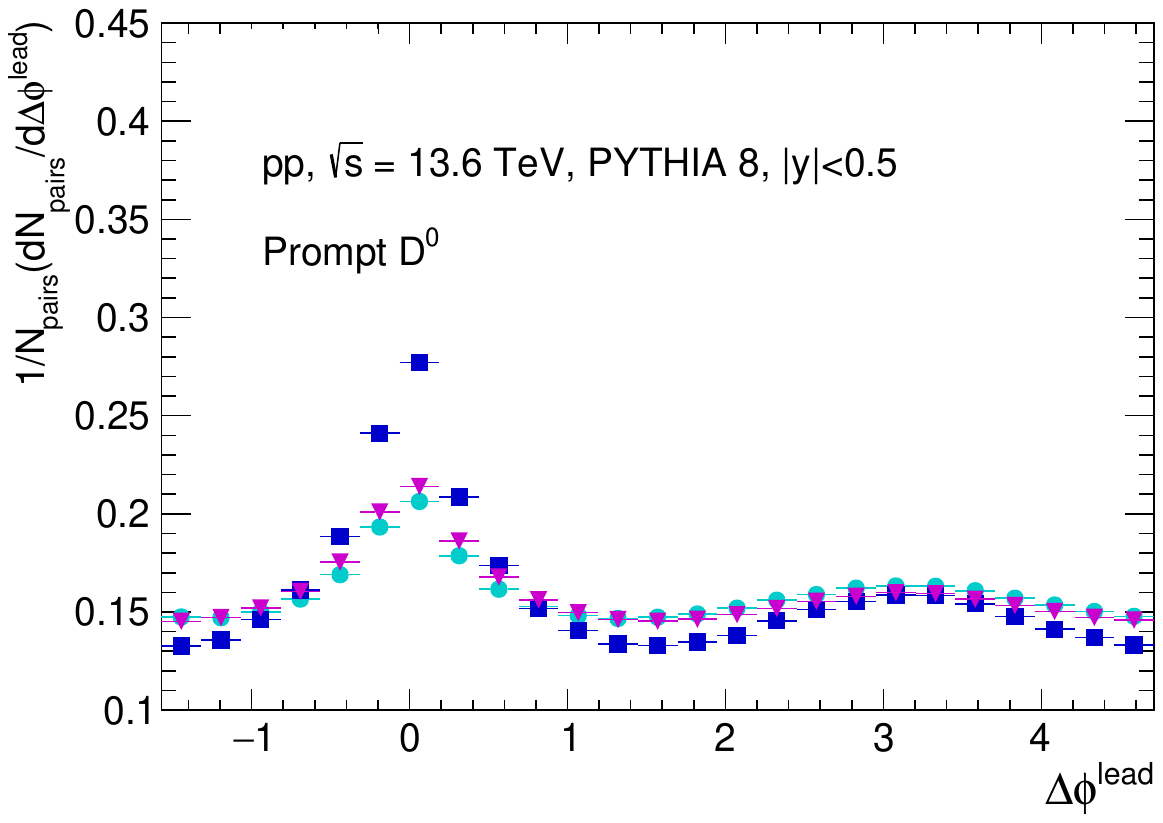}
\includegraphics[scale = 0.4]{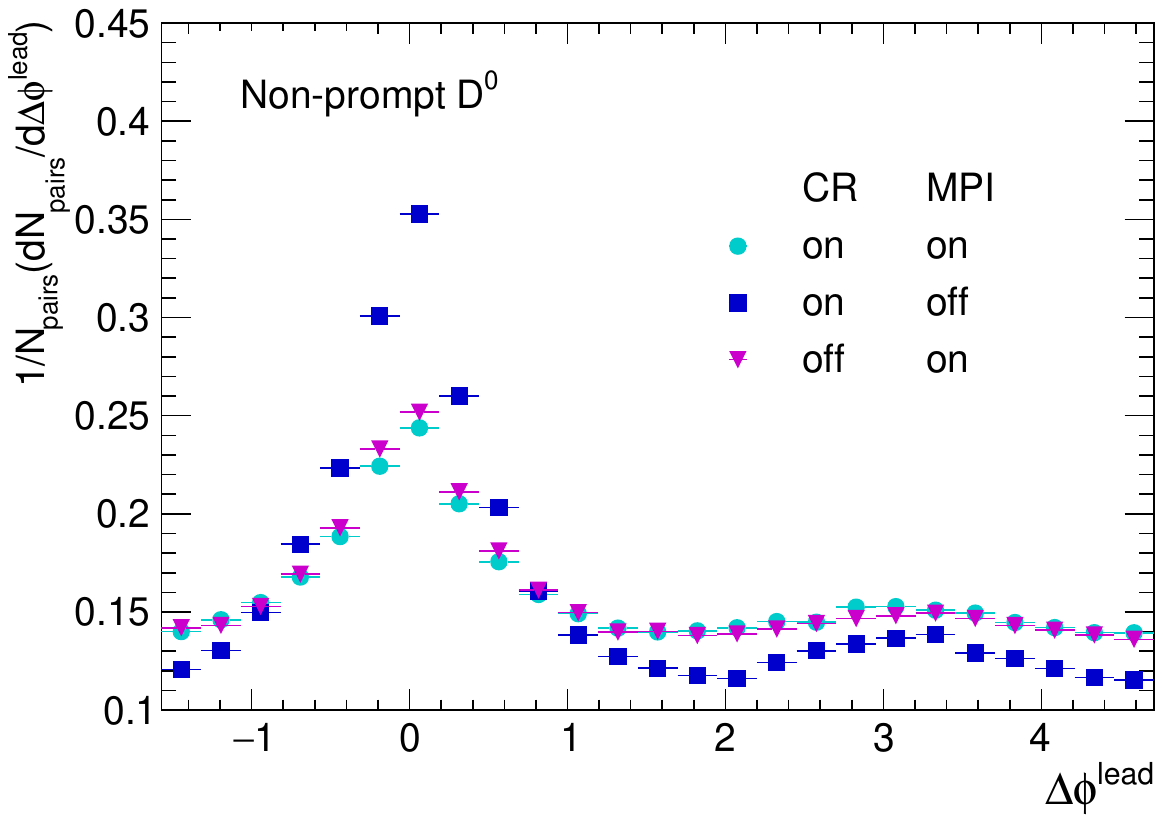}
\caption{$\Delta\phi^{\rm lead}$ distribution between leading charged particle and leading prompt $D^{0}$ (left) and non-prompt $D^{0}$ (right) plotted for the cases--- CR and MPI both switched on, CR on and MPI off, CR and MPI both switched off--- in pp collisions at $\sqrt{s} = 13.6$ TeV using PYTHIA8.} 
\label{fig:deltaphi}
\end{center}
\end{figure*}

Figure~\ref{fig:pTD0leadvsD0region} shows the self-normalised yield for charged particles measured at mid-rapidity ($|\eta|<0.8$) as a function of transverse momentum of the leading prompt (left) and non-prompt (right) $D^{0}$ mesons ($p_{\rm T}^{D^{0}-\rm lead}$) with MPI switched on and off. For the prompt $D^0$ case (left panel), with an increase in $p_{\rm T}^{D^0-\rm lead}$, the relative yield of charged particles increases in all the regions. For the MPI-off case, the relative yield of charged particles increases almost linearly with an increase in prompt-$p_{\rm T}^{D^{0}-\rm lead}$. Here, the momentum of the leading prompt $D^0$ strongly depends upon the momentum transfer between the partons, as only one partonic interaction is allowed. Therefore, the relative yield is the highest for the away region, which is mostly affected by the fragmentation of the jet produced in the opposite direction to that of the $D^0$ meson. This becomes evident as one moves towards events with large values of $p_{\rm T}^{D^{0}-\rm lead}$. In contrast, the slope of this increment of relative yield in the transverse regions is weaker, which hints at the contributions from large jet fragmentation that occurred in the toward and/or away regions with respect to the leading $D^{0}$ particle. The non-alignment of the leading $D^0$ with the leading jet can also induce a finite contribution of fragmented charged particles in the transverse region, which is shown in Fig.~\ref{fig:deltaphi}. 
A similar behaviour is observed in the right panel of Fig.~\ref{fig:pTD0leadvsD0region} for the MPI-off case.

However, for the events with MPI-on, the relative charged particle yield is considerably higher in the transverse regions in comparison to their counterparts in the MPI off scenario, and acquires a saturation beyond $p_{\rm T}^{D^{0}-\rm lead}>10$~GeV/$c$. This is indicative of the dominant MPI activity which increases with $p_{\rm T}^{D^{0}-\rm lead}$, as shown in Fig.~\ref{fig:pTd0leadvspThat}, and saturates at large $p_{\rm T}^{D^{0}-\rm lead}$. For the toward and away regions, the slope of the relative production of charged particles becomes weaker towards the larger $p_{\rm T}^{D^{0}-\rm lead}$ events and starts to saturate, indicating the saturation of $\langle N_{\rm mpi}\rangle$ in addition to contributions from fragmentation processes. The contribution from the fragmentation for the MPI-on case is more prominent in the right panel of Fig.~\ref{fig:pTD0leadvsD0region}, where the relative yield of charged particles is shown as a function of nonprompt $p_{\rm T}^{D^{0}-\rm lead}$. Here, for all three regions, the contribution from underlying MPI cancels out in the relative yield of charged particles and only the effect of fragmentation remains, showing a slower rise in the relative yield of charged particles with an increase in non-prompt $p_{\rm T}^{D^{0}-\rm lead}$.

 \begin{figure}[ht!]
\begin{center}
\includegraphics[scale = 0.44]{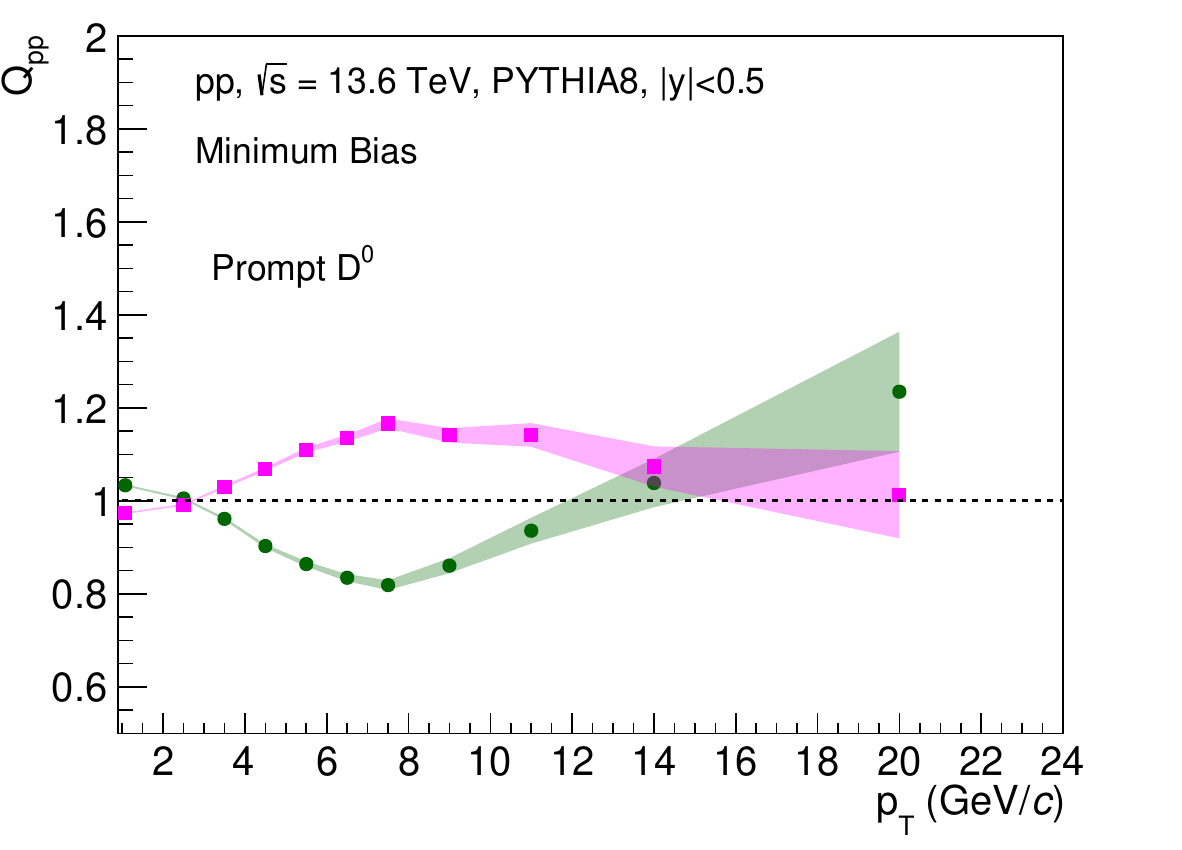}
\includegraphics[scale = 0.44]{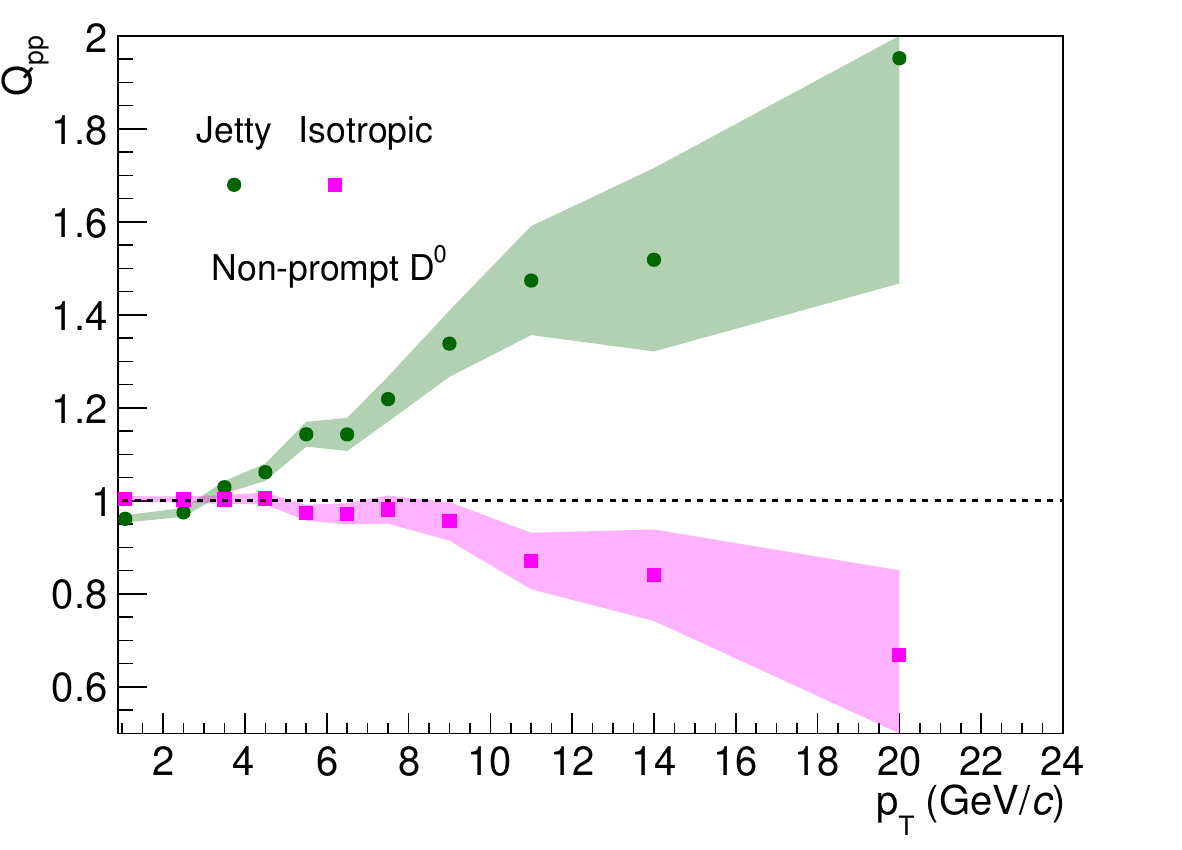}
\caption{$Q_{\rm pp}$ as function of $p_{\rm T}$ for prompt (top) and non-prompt (bottom) $D^{0}$ mesons in jetty and isotropic events of (0--100$\%$) FT0M class in pp collisions at $\sqrt{s} = 13.6$ TeV using PYTHIA8 with MPI and CR switched on. Error bands incorporate the estimated statistical uncertainties.} 
\label{fig:S0D0vsQpp}
\end{center}
\end{figure}

To understand the production mechanisms of prompt and non-prompt $D^0$  with respect to the initial hardest scattering, we show the relative azimuthal dependence of the correlation between the leading charged particle with transverse momenta, $p_{\rm T}^{\rm leading}$, and the leading prompt and non-prompt $D^0$. Figure~\ref{fig:deltaphi} shows the distribution of the relative azimuthal angle between the leading charged particle and the leading $D^0$, i.e., $\Delta\phi^{\rm lead}=\phi_{\rm ch}^{\rm lead}-\phi_{D^{0}}^{\rm lead}$. Here $\phi_{\rm ch}^{\rm lead}$ and $\phi_{D^0}^{\rm lead}$ are the azimuthal angles of the leading charged particle and leading $D^0$, respectively. The left and right panels show the $\Delta\phi^{\rm lead}$ distribution, which is calculated with the azimuthal angles of leading prompt and non-prompt $D^0$, respectively. The cases are shown for different tunes of CR and MPI. As expected, a strong peak is observed near $\Delta\phi^{\rm lead}=0$ (near side) while the observed peak is small and weak near $\Delta\phi^{\rm lead}=\pi$ (away side). This shows that most of the hadrons originating from charm (left plot) and beauty (right plot) quarks are somehow correlated with leading particle production and thus related to the initial hardest scattering. Another interesting thing is that the away-side peak is more visible for the prompt $D^0$ case and is nearly flat for the non-prompt $D^0$ case, which suggests that the charm is produced in the opposite direction to the hard-scattered jet, which is less likely for the beauty quark production. As expected, the near-side peak for both prompt and non-prompt $D^0$ is largest for the MPI-off case. In contrast, both cases with CR-off and CR-on for the MPI-on case are quantitatively similar, where the near side peak is higher for the CR-off case, and the away side peak is higher for the CR-on case for both prompt and non-prompt $D^0$. The near size azimuthal correlation between the leading charged particle and the leading $D^0$ weakens for the environments with large $N_{\rm mpi}$, which becomes stronger for the CR-on case, where partons from independent partonic interactions are made to connect.

Studying the transverse momentum distributions of prompt and non-prompt $D^{0}$ as a function of transverse spherocity can help in identifying the modifications to their spectral shapes in events dominated by hard scatterings, in contrast to the softer isotropic emissions. For this, we depend on the partonic modification factor, $Q_{\rm pp}$, whose construction is motivated by the nuclear modification factor ($R_{\rm AA}$)~\cite{ALICE:2019dfi, Ortiz:2020rwg}. $R_{\rm AA}$ quantifies the changes to the particle spectra induced by nuclear matter in heavy-ion collisions by comparing them to a pp baseline scaled by the number of binary nucleon-nucleon interactions. $Q_{\rm pp}$ performs an analogous job in pp collisions, which quantifies the modification of spectral shape in a set of events selected based on a specific event property in comparison to the minimum bias events. For this particular study, the $p_{\rm T}$-spectra in a specific $S_{0}$-event class is scaled by the average number of $D^{0}$ mesons in that class, and then compared to the $S_{0}-$integrated events in the denominator as shown below:

\begin{equation}
Q_{\rm pp} = \frac{d^2N_{\rm D^{0}}^{S_{\rm 0}}/\langle N_{\rm D^{0}}^{S_{\rm 0}} \rangle d\eta dp_{\rm T}} {d^2N_{\rm D^{0}}^{\rm S_{0}-Intg.}/\langle N_{\rm D^{0}}^{\rm S_{0}-Intg.} \rangle d\eta dp_{\rm T}}
\label{eq:Qpp}
\end{equation}
 where $\langle N_{\rm D^{0}}^{\rm S_{0}-Intg.} \rangle / \langle N_{\rm D^{0}}^{S_{\rm 0}} \rangle$ is used as normalization constant similar to $1/\langle N_{\rm coll} \rangle$ in $R_{\rm AA}$.

Figure~\ref{fig:S0D0vsQpp} presents $Q_{\rm pp}$ as a function of $p_{\rm T}$ for prompt (upper plot) and non-prompt (lower plot) $D^{0}$ mesons in $|y|< 0.5$ for jetty and isotropic event selections of (0--100\%) FT0M class in pp collisions at $\sqrt{s}=13.6$~TeV using PYTHIA8. As shown in Fig.~\ref{fig:pTD0leadvsS0}, the fact that $S_{0}$ and $N_{\rm mpi}$ are positively correlated is well established. It is also understood from Fig.~\ref{fig:D0vsNmpi} that the heavy-flavour particle production is significantly favoured in high $N_{\rm MPI}$ events, hence in isotropic events. In line with these findings and consistent to the observations for midrapidity prompt $J/\psi$ in Ref.~\cite{Radhakrishnan:2025owp}, $Q_{\rm pp}$ for prompt $D^{0}$ from isotropic events is gently seen to rise with increasing $p_{\rm T}$ while the corresponding trend of $Q_{\rm pp}$ for jetty event class shows a slightly steeper fall. The two curves cross each other at $\approx 2.5$~GeV/$c$ and continue their trend till $p_{\rm T}$ $\approx 8$~GeV/$c$. Beyond this region, $Q_{\rm pp}$ from isotropic (jetty) events falls (rises) back to approach unity. The very (rising and falling) behaviour of the ratio $Q_{\rm pp}$ is indicative of the hardening and softening of the $D^{0}$ $p_{\rm T}$ spectra in isotropic and jetty events with respect to the integrated spherocity events~\cite{Radhakrishnan:2025owp}. 

As can be realised from Fig.~\ref{fig:pTD0leadvsS0} and Table~\ref{Table:1}, the behaviour for $Q_{\rm pp}$ upto $p_{\rm T}\approx 7-8$~GeV/$c$ for prompt $D^{0}$ is consistent with the isotropic events being associated with larger average $N_{\rm mpi}$, which enhances the relative contribution of prompt $D^{0}$ production at low and intermediate transverse momenta. This is reflected directly in the upper plot of Fig.~\ref{fig:S0D0vsQpp}. However, this relative dominance of isotropic events begins to diminish as we shift from the softer charm-hadron production region into the higher $p_{\rm T}$ regime. While increased $N_{\rm mpi}$ enhances charm production at low and intermediate $p_{\rm T}$, it does not significantly increase the probability of very high-$p_{\rm T}$ charm production, which is dominated by the hardest partonic scatterings. As a result, isotropic events gradually lose their relative advantage at high-$p_{\rm T}$, and jetty events start to dominate. High $p_{\rm T}$  prompt $D^{0}$ mesons predominantly originate from the hardest 2$\rightarrow$2 partonic scatterings, which naturally correspond to jetty event topologies. This explains the rise of $Q_{\rm pp}$ for jetty events beyond $p_{\rm T}>$ 15~GeV/$c$. On the contrary, although MPI can accompany a hard scattering, the dominant contribution to b-hadron production comes from the primary high-$\hat{p}_{\rm T}$ 2$\rightarrow$2 process itself. This naturally biases such events toward the lowest spherocity (\textit{i.e.} jetty) classes.

\begin{figure*}[ht!]
\begin{center}
\includegraphics[scale = 0.92]{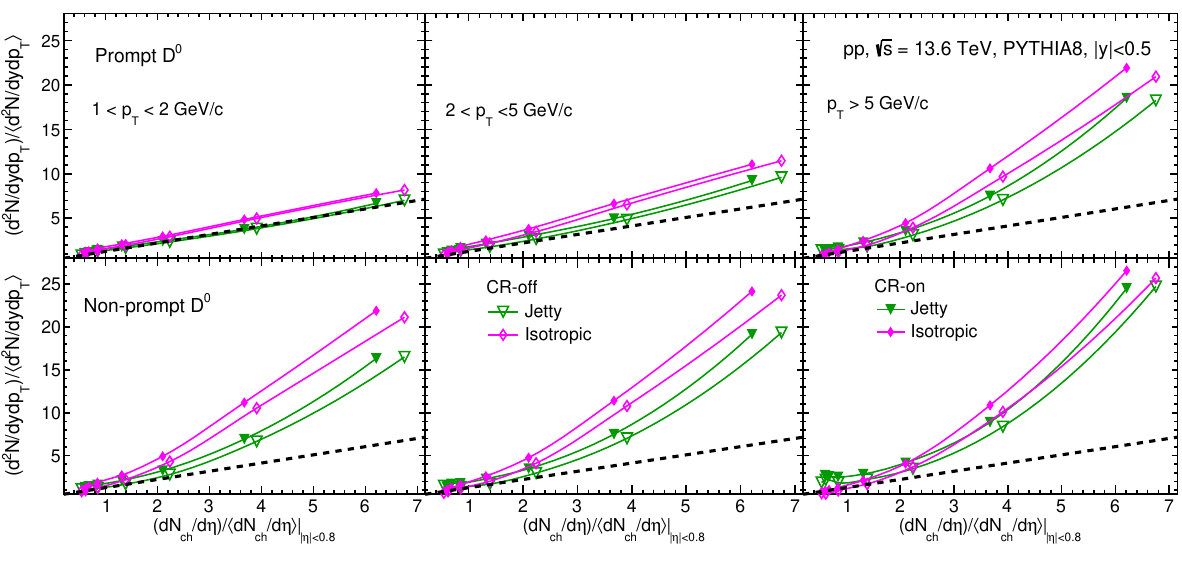}
\caption{Self-normalised prompt (top) and non-prompt (bottom) $D^{0}$ meson yield at midrapidity ($|y|<0.5$) as a function of normalized mid-rapidity charged-particle multiplicity density for low-$p_{\rm T}$ region (left), intermediate-$p_{\rm T}$ region (middle), high-$p_{\rm T}$ region (right) in jetty and isotropic events for CR on and off cases in pp collisions at $\sqrt{s} = 13.6$ TeV using PYTHIA8 with MPI switched on.}
\label{fig:SNYieldvsS0CR}
\end{center}
\end{figure*}

To analyse the transverse momentum and multiplicity dependence of $D^{0}$ production, their self-normalised yield is studied in experiments in various $p_{\rm T}$ intervals and multiplicity bins, where a stronger-than-linear trend is observed not only for $D^{0}$ mesons but also for $J/\psi$ and strange hadrons~\cite{ALICE:2020msa,ALICE:2021zkd,ALICE:2025fzz,ALICE:2019avo}. Such a behaviour is generally attributed to the autocorrelation bias that arises as a result of performing event classification as well as estimation of the observable of interest using the charged particles in the same phase space. Fig.~\ref{fig:SNYieldvsS0CR} presents the self-normalised yield for prompt (upper panels) and non-prompt (lower panels) $D^{0}$ meson at midrapidity ($|y|~<~0.5$) in different $p_{\rm T}$ intervals --- $1<p_{\rm T}<2$~GeV/$c$ (left), $2<p_{\rm T}<5$~GeV/$c$ (middle) and $p_{\rm T}>5$~GeV/$c$ (right)--- as a function of self-normalised charged-particle multiplicity density for jetty and isotropic events with CR cases ``on" and ``off" for the (0--100$\%$) FT0M class of pp collisions at $\sqrt{s} = 13.6$~TeV using PYTHIA8. The selections of $D^{0}$ and charged-particle multiplicity are done in different rapidity regions, thereby reducing the possible autocorrelation effects, while biases due to $D^{0}$ and $S_{0}$ estimation being in the same rapidity ought to result in self-correlation bias~\cite{Radhakrishnan:2025owp}. 

As anticipated from the results of pp collisions at $\sqrt{s}=13$~TeV~\cite{Bailung:2021jow}, the $p_{\rm T}$-differential self-normalised yield of $D^{0}$ mesons increases stronger-than-linearly with increasing charged-particle multiplicity, and the increment is seen to grow steeper as $p_{\rm T}$ increases. In fact, the highest multiplicity pp events correspond to scenarios with a large overlap of strings and enhanced MPI activity, which in turn increases the overall probability of charm production. The steeper rise observed at higher $p_{\rm T}$ arises from an event-selection bias, whereby the production of higher $p_{\rm T}$ $D^{0}$-mesons preferentially selects more central, high-multiplicity collisions. Additionally, in comparison to prompt $D^{0}$, the deviation from linearity for the self-normalised yield of non-prompt $D^{0}$ is observed to start earlier and attain higher magnitudes throughout the $p_{\rm T}$ intervals studied, once again reminding us of the important role of higher $N_{\rm mpi}$ in increasing the relative yield of non-prompt $D^{0}$ (as evident from Fig.~\ref{fig:D0vsNmpi}). Moreover, the relative prompt and non-prompt $D^{0}$ yield is found to be enhanced in isotropic events than in jetty events due to the well-established correlation between $D^{0}$ yield and the $N_{\rm mpi}$. Lastly, if one looks at the effect of color reconnection in Fig.~\ref{fig:SNYieldvsS0CR}, it is understood that the CR has an influence only/mostly on the final-state multiplicity, and has little effect on the heavy-flavour production. As already described elsewhere, with CR on, strings from different partonic scatterings reconnect and shorten, leading to fewer charged particles, whereas in the CR-off case, strings fragment independently, producing a larger multiplicity for the same underlying MPI activity. As a result, the same multiplicity percentiles correspond to a higher $N_{\rm ch}$ in the CR-off case and a slightly lower one in the CR-on case (see Table~\ref{Table:1}). Moreover, $D^{0}$ production scales with MPI rather than the visible multiplicity, hence stands unaffected by CR, which happens at a later stage. Consequently, the only difference reflected between the CR-off and CR-on cases in Fig.~\ref{fig:SNYieldvsS0CR} is a shift in the value of multiplicity in the horizontal direction, where the degree of shift increases with increasing non-linearity of the curve.

\begin{figure}[ht!]
\begin{center}
\includegraphics[scale = 0.42]{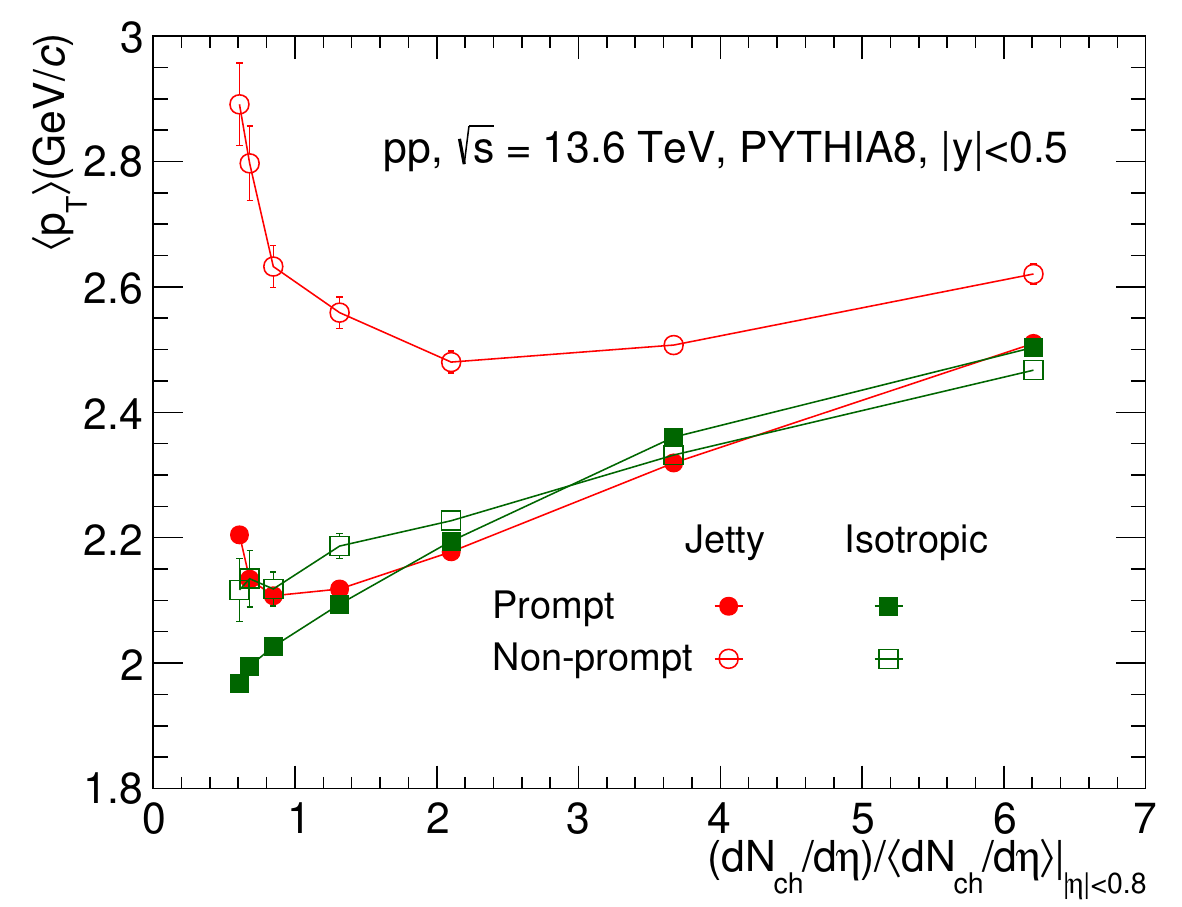}
\caption{Mean transverse momentum ($\langle p_{\rm T}\rangle$) of prompt and non-prompt $D^{0}$ mesons as a function of self-normalised mid-rapidity charged particle multiplicity density measured for jetty and isotropic events in pp collisions at $\sqrt{s}~=~13.6$~TeV using PYTHIA8.}
\label{fig:meanpT}
\end{center}
\end{figure}

\begin{figure}[ht!]
\begin{center}
\includegraphics[scale = 0.44]{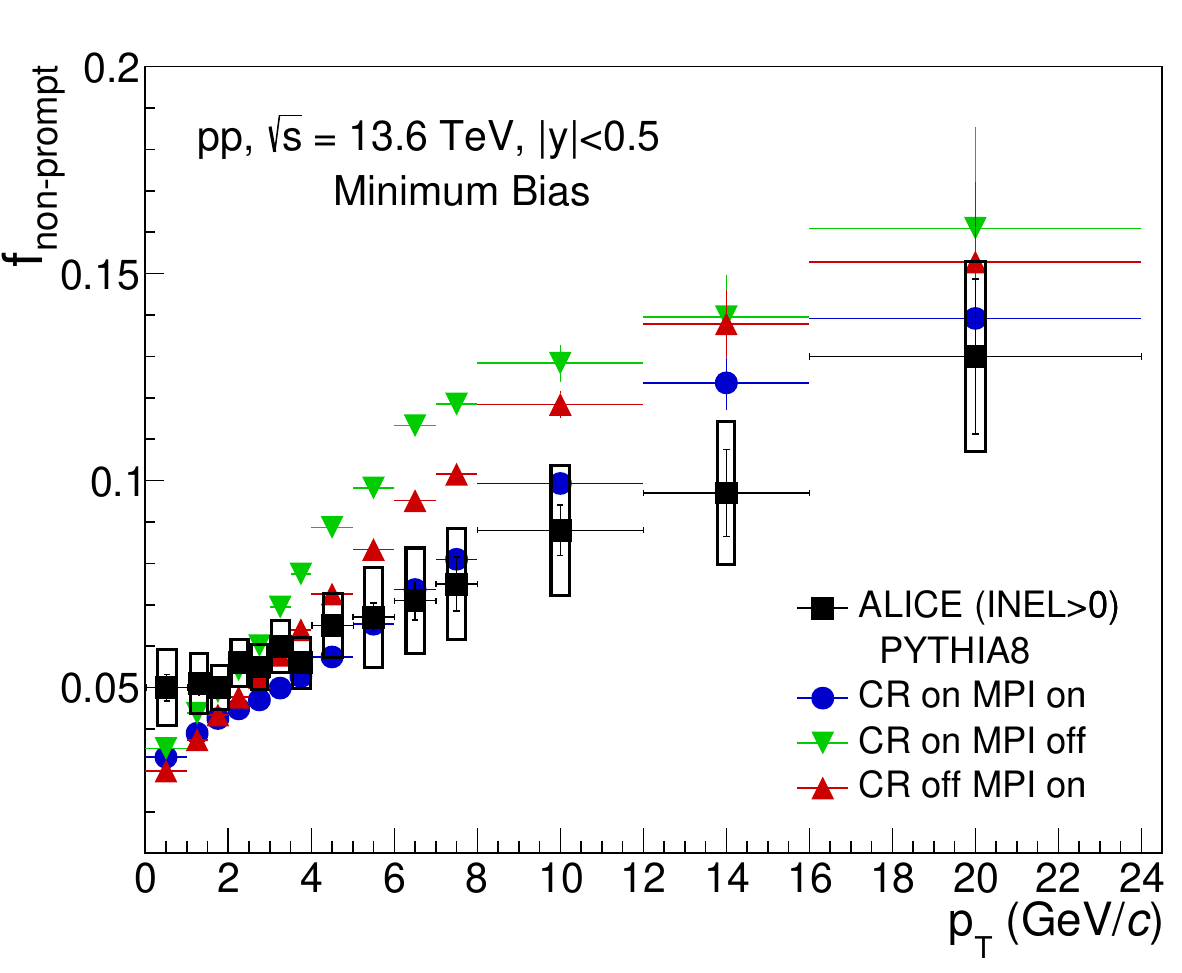}
\caption{Comparison of PYTHIA8 generated data for the cases, MPI-on CR-on, MPI-on CR-off, and MPI-off CR-on, with ALICE~\cite{Garcia:2025xnf} results for non-prompt $D^{0}$ fraction as a function of transverse momentum in pp collisions at $\sqrt{s}~=~13.6$~TeV.}
\label{fig:fBALICE}
\end{center}
\end{figure}

\begin{figure*}[ht!]
\begin{center}
\includegraphics[scale = 0.42]{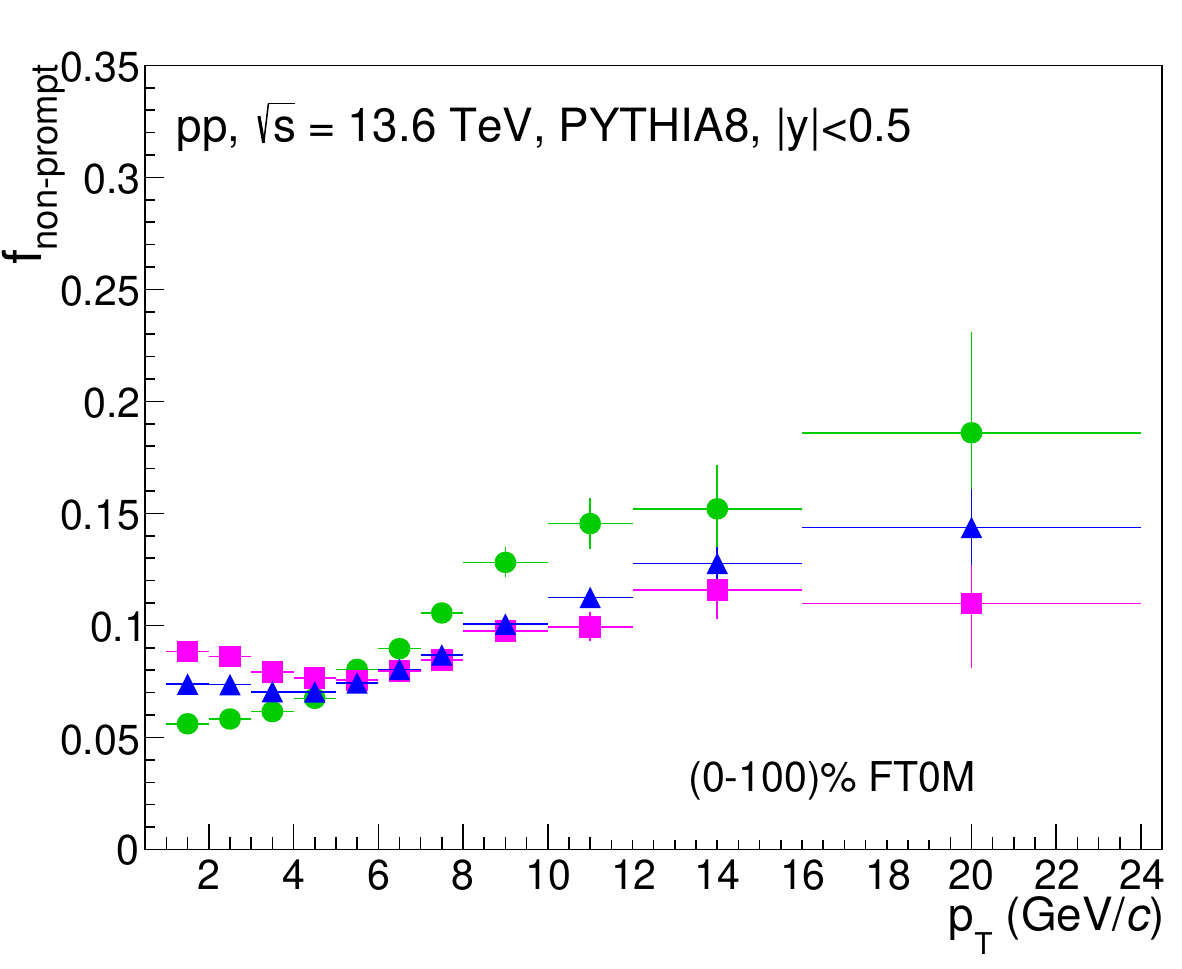}
\includegraphics[scale = 0.42]{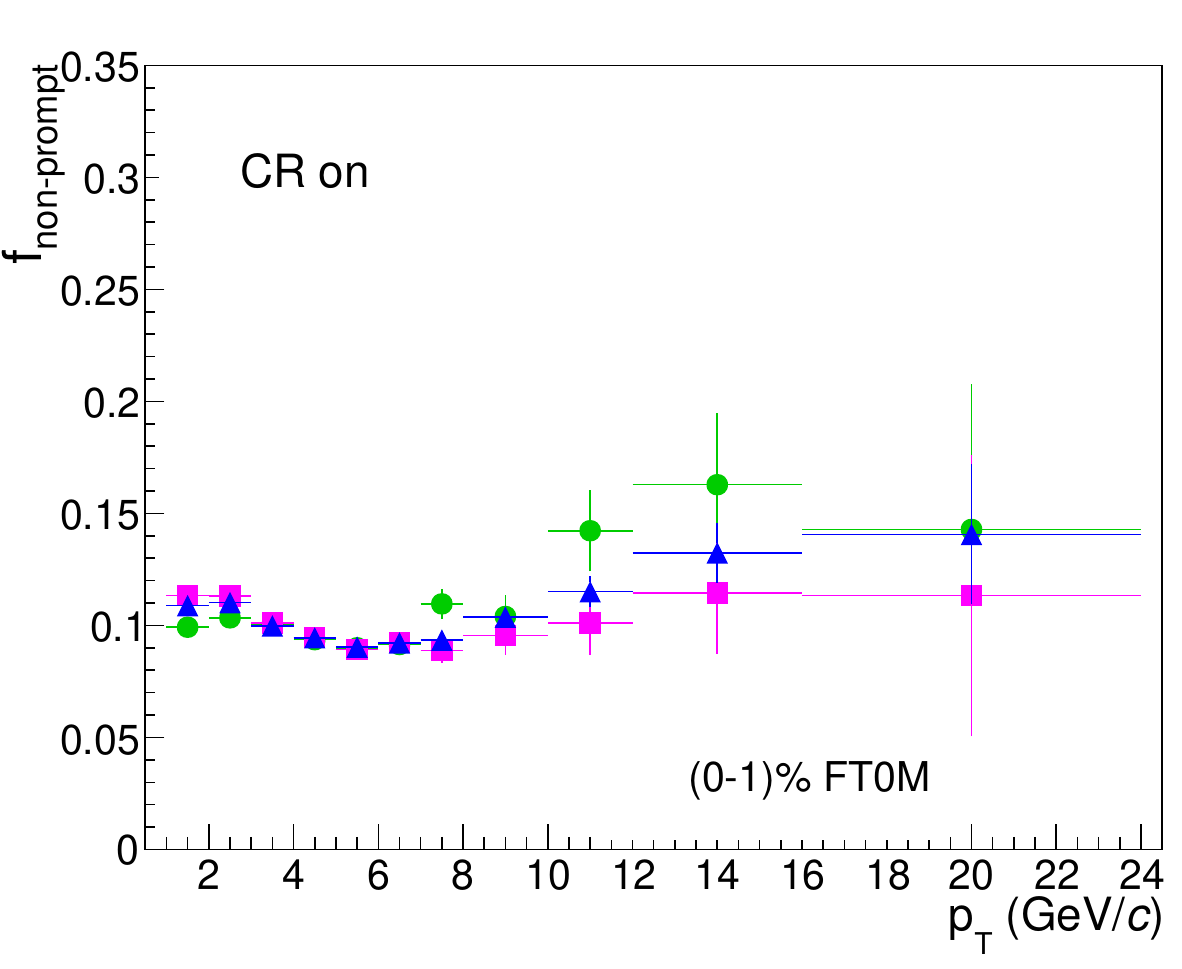}
\includegraphics[scale = 0.42]{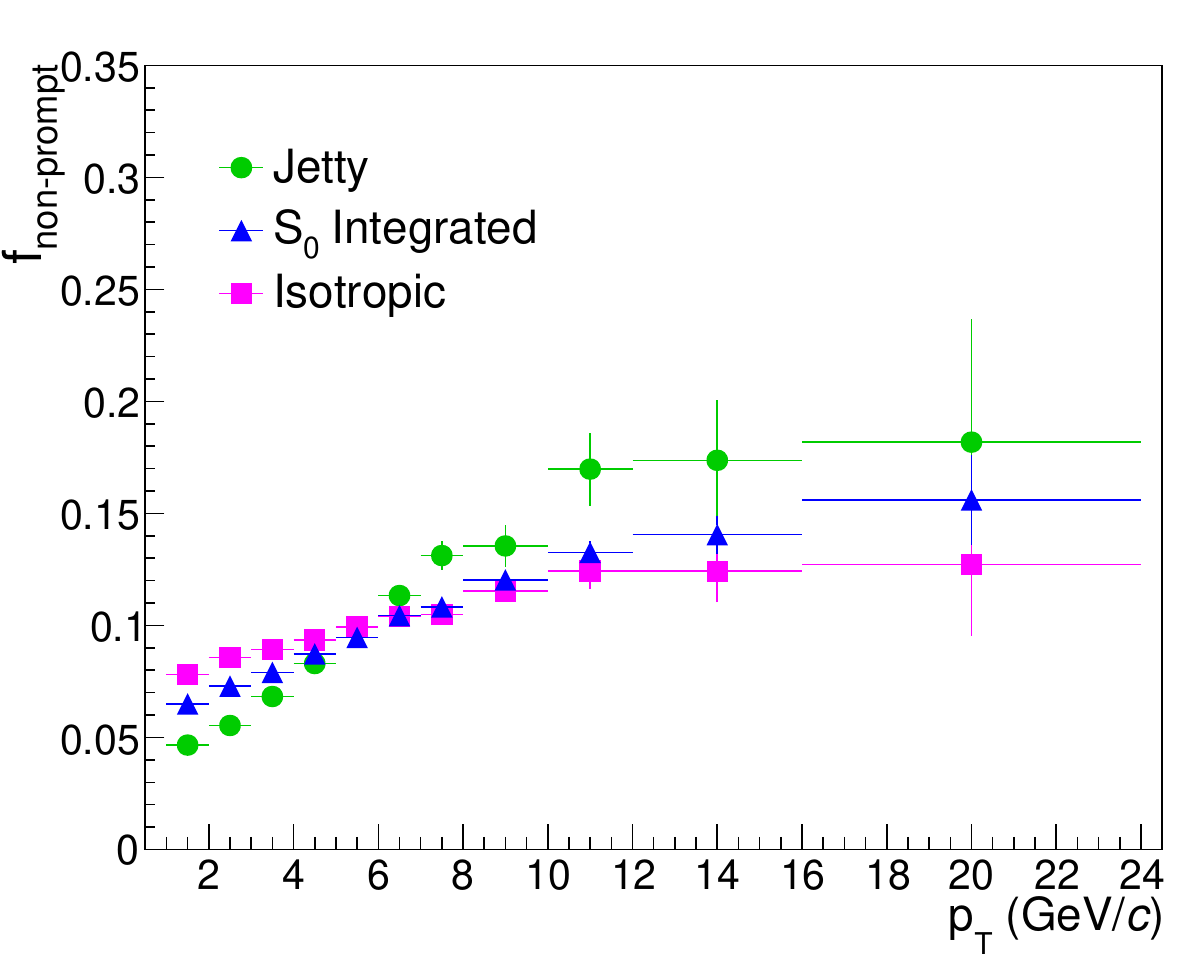}
\includegraphics[scale = 0.42]{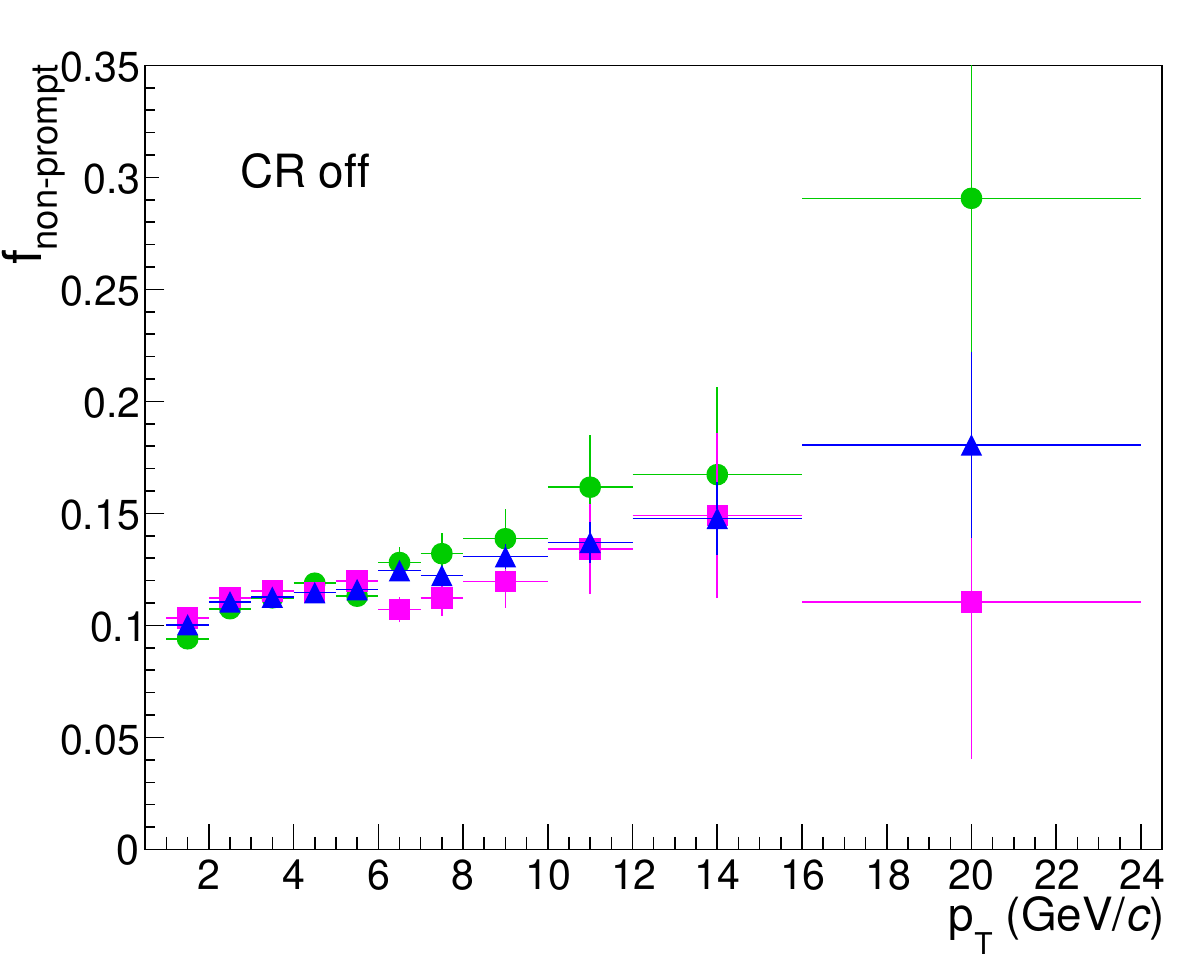}
\caption{Ratio of differential yield of non-prompt $D^{0}$ to prompt $D^{0}$ for $(0-100)\%$ (left) and $(0-1)\%$ (right) FT0 multipicity classes for jetty and isotropic events, for CR-on (upper) and CR-off (lower) cases in pp collisions at $\sqrt{s}~=~13.6$~TeV using PYTHIA8 with MPI switched on.}
\label{fig:fracNonpromtp}
\end{center}
\end{figure*}

Figure \ref{fig:meanpT} shows the correlation of average transverse momentum ($\langle p_{\rm T}\rangle$) of prompt and non-prompt $D^{0}$ mesons from jetty and isotropic events to that of the normalised mid-rapidity charged-particle multiplicity density. One clear observation that can be made is the dominance of $\langle p_{\rm T}\rangle$ of non-prompt $D^{0}$ over prompt $D^{0}$, throughout the multiplicity classes for jetty events and in the low-multiplicity range for isotropic events. Also, $\langle p_{\rm T}\rangle$ of non-prompt $D^{0}$s from jetty events is steadily greater than that from the isotropic events, throughout the multiplicity range, while such a dominance is found for the prompt $D^{0}$s from jetty events only in the lower multiplicity classes. Additionally, a sharp (slow) increase of $\langle p_{\rm T}\rangle$ of non-prompt (prompt) $D^{0}$ towards the lowest multiplicity jetty events is also observed. These results draw us to the following conclusions: 
\begin{enumerate}
    \item Except for the low multiplicity events, an increase in multiplicity is seen to be accompanied by a gradual increase in the mean transverse momentum of the produced $D^{0}$ mesons in those events. As discussed before, an increase in multiplicity is associated with an increase in $N_{\rm mpi}$ and a decrease in the impact parameter of collisions. A smaller impact parameter allows a larger number of violent interactions that can produce $D^{0}$ with larger $p_{\rm T}$, as illustrated in Fig.~\ref{fig:pTd0leadvspThat}.

    \item The hardest 2$\rightarrow$2 partonic scatterings that produce non-prompt or prompt $D^{0}$ are usually single hard scatterings, often associated with low-multiplicity, jet-dominated topologies. The characteristic back-to-back dijet structure of such events is identified by the event classifier, transverse spherocity, as jetty events. In such jetty-identified events, the dominance of hard-scattering products and the reduced contribution from soft particle production lead to a larger event-averaged transverse momentum, resulting in the highest values of $\langle p_{\rm T}\rangle$, particularly for the non-prompt $D^{0}$ case.
    
    \item $\langle p_{\rm T}\rangle$ of non-prompt $D^{0}$s are higher than that of the prompt ones --- as non-prompt $D^{0}$ mesons inherit a substantial fraction of the $p_{\rm T}$ of their parent beauty hadrons whose typical $p_{\rm T}$ is larger than that of promptly produced charm hadrons. This is referred to as the characteristic mass ordering of $\langle p_{\rm T}\rangle$, which has been observed in studies relating to both light and heavy-flavour hadrons~\cite{ALICE:2019hno, Radhakrishnan:2025owp}. However, as we approach the highest multiplicity class, the $\langle p_{\rm T}\rangle$ of the prompt $D^{0}$s from isotropic events becomes comparable to, or slightly exceeds, that of non-prompt $D^{0}$ mesons. This reflects the increased contribution of softer beauty production in high-activity isotropic events, while prompt charm production continues to benefit from the enhanced event activity.
    
    \item The dependence of $\langle p_{\rm T}\rangle$ on transverse spherocity $S_{0}$ weakens with increasing charged-particle multiplicity. At high multiplicities, the $\langle p_{\rm T}\rangle$ values for prompt and non-prompt $D^{0}$ mesons from different event-shape classes converge, indicating a reduced sensitivity to event topology. An exception is observed for non-prompt $D^{0}$ mesons in jetty events, which continue to exhibit a slightly higher $\langle p_{\rm T}\rangle$ due to the dominance of hard-scattering contributions.
\end{enumerate}

To further disentangle the relative production mechanisms of charm and beauty hadron in pp collisions, the fraction of $D^{0}$ mesons originating from b-decay to that of prompt $D^{0}$, denoted as $f_{\rm non-prompt}$, can be studied as a function of $p_{\rm T}$ for various event topologies. This observable can guide us in understanding the interplay between hard-partonic scatterings that predominantly govern beauty production and the MPI-driven processes that contribute to prompt charm production, thus helping in probing how charm and beauty respond differently to the underlying event activity and event shape in small collisions. We begin by comparing PYTHIA8 results for MB events with the corresponding data from ALICE for pp collisions at $\sqrt{s}=13.6$~TeV, as presented in Fig.~\ref{fig:fBALICE}. $f_{\rm non-prompt}$ from PYTHIA8 is found to underestimate the ALICE results in the low $p_{T}$ region while overestimating it in the high $p_{T}$ regime. However, one can also observe that the non-prompt to prompt $D^{0}$ ratio obtained using PYTHIA for the MPI-on CR-on case agrees qualitatively well and explains the ALICE data much better than the corresponding MPI-on CR-off and MPI-off CR-on results from PYTHIA8.

$f_{\rm non-prompt}$ as a function of $p_{\rm T}$ for jetty, $S_{0}$-integrated and isotropic events in (0--100$\%$) and (0--1$\%$) FT0M classes is presented in the left and right columns of Fig.~\ref{fig:fracNonpromtp} respectively. The upper row plots represent the MPI-on--CR-on scenario, while the lower row plots present the MPI-on--CR-off case. The magnitude of the $f_{\rm non-prompt}$ ratio, as expected from previous studies, confirms that the prompt $D^{0}$ yield is significantly larger than the non-prompt $D^{0}$s produced, owing to the higher production threshold and larger mass of the beauty quark compared to the charm quark~\cite{Weber:2018ddv, Radhakrishnan:2025owp}. A significant dependence of $f_{\rm non-prompt}$ on both $p_{\rm T}$ as well as $S_{0}$ is also observed. 

\begin{itemize}
\item \textbf{CR-on} case:  In the highest multiplicity pp events (right plot), as we traverse from low to intermediate $p_{\rm T}$ region, $f_{\rm non-prompt}$ starts to drop, for all $S_{0}$ classes, though a slight dominance of isotropic events is observed over jetty events. This gradually falling trend with an increase in $p_{\rm T}$ might be indicative of the relatively stronger increase of prompt $D^{0}$ production in this region. However, after $p_{\rm T}>5$~GeV/$c$, $f_{\rm non-prompt}$ starts to rise strongly for jetty events while the trend from the isotropic case flattens out. On the other hand, for the minimum-bias (MB) case (left plot), a very clear distinction between all three $S_{0}$ classes persists over the full $p_{\rm T}$ range. In contrast to the dip-and-rise pattern seen for all $S_{0}$ classes in the (0–1)$\%$ FT0M selection, the MB results show such a feature only for the isotropic class. The rise of the fraction $f_{\rm non-prompt}$ towards higher $p_{\rm T}$ represents the hardening of transverse momentum spectra of beauty-hadrons with respect to the charmed hadrons~\cite{Radhakrishnan:2025owp}. Consistent with earlier observations, the $p_{\rm T}$-spectra of non-prompt $D^{0}$ mesons in jetty events are harder at higher $p_{\rm T}$ than those in isotropic ones, emphasising the dominant role of hard partonic scatterings in jetty topologies. 

\item \textbf{CR-off} case: As the color reconnection is switched off, the very dip-like structure earlier observed for $f_{\rm non-prompt}$ in the $4<p_{\rm T}<8$~GeV/$c$ region no more appears. Instead, $f_{\rm non-prompt}$ suffers a steady increase with $p_{\rm T}$, where a clear distinction between $S_{0}$ classes is observed for the (0--100)$\%$ FT0M class--- like in the MPI-on--CR-on case for MB pp events, the low $p_{\rm T}$ non-prompt $D^{0}$ production is more favoured in isotropic events while the higher-$p_{\rm T}$ non-prompt $D^{0}$ predominantly originates in jetty events. On the other hand, $S_{0}$-dependence of $f_{\rm non-prompt}$ is found to be little, as one shifts the focus to the highest multiplicity class.
\end{itemize}
This comparative study highlights the important role of CR in enhancing the prompt $D^{0}$ production in the intermediate $p_{\rm T}$ region and the requirement of higher transverse momentum for non-prompt $D^{0}$ production. This also hints at the inherent isotropicity in high-multiplicity pp events, as the $f_{\rm non-prompt}$ trend of (0--1)$\%$ FT0M class is found to be qualitatively similar to that of isotropic events of MB class.

\section{Summary} 
\label{sec:summary}


This work provides an in-depth analysis of the production of prompt and non-prompt $D^0$ mesons with emphasis on their dependence on $\hat{p}_{\rm T}$, MPI, and CR mechanisms within the framework of PYTHIA8. The experimentally measurable quantities, such as charged particle multiplicity, transverse spherocity, $p_{\rm T}^{D^{0}-\rm lead}$, act as a proxy for MPI and $\hat{p}_{\rm T}$. The study is performed in pp collisions at $\sqrt{s}=13.6$ TeV, and comparison with experimental measurements is provided, wherever possible. The point-to-point summary of the work is provided as follows.

\begin{enumerate}
    \item  The self-normalized yield of prompt and non-prompt $D^{0}$ mesons increase with both $\hat{p}_{\rm T}$ and $N_{\rm mpi}$. Non-prompt $D^{0}$ is found to be exceptionally sensitive to $\hat{p}_{\rm T}$ and $N_{\rm mpi}$, hinting at the dominant role of early hard scatterings in beauty production.
    
    \item The strong correlation of $p_{\rm T}$ of the leading $D^{0}$ meson with $\hat{p}_{\rm T}$ and $N_{\rm mpi}$ validates $p_{\rm T}^{D^{0}-\rm lead}$ as an experimentally accessible proxy for the transverse momentum transfer in the hardest partonic scattering.

    \item Charged-particle production in the transverse region is strongly enhanced in MPI-on events, confirming the significant contribution of MPI to the underlying event, while toward and away regions are increasingly dominated by prompt $D^{0}$ production through both jet fragmentation and MPI activity.
    
    \item Prompt $D^{0}$ mesons exhibit an enhancement in isotropic events at low and intermediate $p_{\rm T}$, while transitioning to jet-dominated production at high $p_{\rm T}$. 
    
    \item A stronger-than-linear increase of self-normalized yield with multiplicity is observed for prompt $D^{0}$ and more strongly for non-prompt $D^{0}$, with the rate of increment being the largest for the highest $p_{\rm T}$ $D^{0}$-- indicates how strongly sensitive the beauty-production is to event activity.
    
    \item With event activity, $\langle p_{\rm T}\rangle$ of both prompt and non-prompt $D^{0}$ mesons increases,  where non-prompt $D^{0}$ from jetty events consistently exhibits higher $\langle p_{\rm T}\rangle$, reflecting the harder beauty hadron kinematics. Sensitivity to event topology is reduced at high multiplicities.
    
    \item $f_{\rm non-prompt}$ ratio reveals a relative enhancement of prompt $D^{0}$ production at low and intermediate $p_{\rm T}$, particularly in isotropic events for the CR-on case and a dominance of hard-scattering driven non-prompt $D^{0}$ production at high $p_{\rm T}$, in jetty events.
\end{enumerate}

With this study, we demonstrated that $\hat{p}_{\rm T}$ and MPI play a consequential role in the production of prompt and non-prompt $D^0$. In contrast, the role of CR is found to be minimally significant. Although the sensitivity of $\hat{p}_{\rm T}$ and MPI varies from one observable to another under study, $S_0$ and $p_{\rm T}^{D^0-\rm lead}$ can be used to reproduce these effects in experiments. Finally, the results obtained in this study once again reaffirm the ability of transverse spherocity to effectively discriminate between soft and hard particle production and demonstrate its applicability in the heavy-flavour sector.

\section*{Acknowledgement}
A.M.K.R. acknowledges the doctoral fellowships from the DST INSPIRE program of the Government of India. The authors gratefully acknowledge the DAE-DST, Government of India, funding under the mega-science project “Indian participation in the ALICE experiment at CERN” bearing Project No. SR/MF/PS-02/2021-IITI(E-37123).


\end{document}